\documentclass[12pt,preprint]{aastex}
\usepackage{graphicx,float,subfigure,amsmath,amsthm,amssymb,multirow,color,upgreek}
\usepackage{soul}
\graphicspath{{./figures/}}

\newcommand{\msun}{\hbox{${M}_{\odot}$}}

\newcommand{\msig}{$M_{bh}-\sigma$}
\newcommand{\vd}{$\sigma$}
\newcommand{\m}{$M_{bh}$}
\newcommand{\ri}{$r_i$}

\newcommand{\be}{\begin{equation}}
\newcommand{\ee}{\end{equation}}
\newcommand{\ba}{\begin{eqnarray}}
\newcommand{\ea}{\end{eqnarray}}

\newcommand{\simgt}{\lower 2pt \hbox{$\, \buildrel {\scriptstyle >}\over {\scriptstyle\sim}\,$}}
\newcommand{\simlt}{\lower 2pt \hbox{$\, \buildrel {\scriptstyle <}\over {\scriptstyle\sim}\,$}}
\newcommand{\ls}{\lower 2pt \hbox{$\;\scriptscriptstyle \buildrel<\over\sim\;$}}
\newcommand{\gs}{\lower 2pt \hbox{$\;\scriptscriptstyle \buildrel>\over\sim\;$}}
\newcommand{\sarc}{$^{\prime\prime}\!\!.$}

\definecolor{Mygrey}{gray}{0.75}

\begin{document}

\def\arcsec{$^{\prime\prime}$}
\def\arcmin{$^{\prime}$}
\def\degr{$^{\circ}$}

\title{A Bayesian Monte-Carlo Analysis of the M-Sigma Relation}
\author{Leah K. Morabito\altaffilmark{1} and Xinyu Dai\altaffilmark{1}}
\altaffiltext{1}{Homer L. Dodge Department of Physics \& Astronomy,
University of Oklahoma, Norman, OK 73019, USA, morabito@nhn.ou.edu, dai@nhn.ou.edu}

\begin{abstract}
We present an analysis of selection biases in the \msig\ relation using Monte-Carlo simulations including the sphere of influence resolution selection bias and a selection bias in the velocity dispersion distribution. We find that the sphere of influence selection bias has a significant effect on the measured slope of the \msig\ relation, modeled as $beta_{intrinsic}=-4.69+2.22\beta_{measured}$, where the measured slope is shallower than the model slope in the parameter range of $\beta > 4$, with larger corrections for steeper model slopes. 
Therefore, when the sphere of influence is used as a criterion to exclude unreliable measurements, it also introduces a selection bias that needs to be modeled to restore the intrinsic slope of the relation.
We find that the selection effect due to the velocity dispersion distribution of the sample, which might not follow the overall distribution of the population, is not important for slopes of $\beta\sim4$--6 of a logarithmically linear $M_{bh}-\sigma$ relation, which could impact some studies that measure low (e.g., $\beta<4$) slopes. 
Combining the selection biases in velocity dispersions and the sphere of influence cut, we find the uncertainty of the slope is larger than the value without modeling these effects, and estimate an intrinsic slope of $\beta=5.28_{-0.55}^{+0.84}$.   
\end{abstract}


\section{Introduction}
The relation between super-massive black hole (SMBH) masses, \m, and their host galaxy velocity dispersion, \vd, is one of the central themes in galaxy and AGN studies. The number of reliable SMBH estimates has more than quadrupled since the early stages of investigation \citep[e.g.,][]{fm00,g00}, with the latest samples containing $\sim64$ galaxies \citep[e.g.,][]{graham11,mc11}. 
Even with an increase in sample size, the relation remains tight, e.g., with a scatter of only 0.43 dex \citep{graham11}, tantalizingly suggestive of a strong link between SMBH and galaxy formation, and indicative of AGN feedback processes. 
Many theoretical predictions and models are based on this relation, describing the interaction between SMBH and the bulge of the galaxy through energy-driven winds \citep[e.g.,][]{silk98} or momentum-driven winds \citep[e.g.,][]{fabian99,k03}. The \msig\ relation has been used to support many theories connecting the central black hole to other host properties of the galaxy \citep[e.g.,][]{ff05,adams03,mh08,bw06,dm08,onb08}, where connections as far as with dark matter halos have also been suggested \citep[e.g.,][]{bs10}. 
There are also studies \citep[e.g.,][]{gultekin09,graham11,beifiori12} that examine the possibility of the \msig\ relation varying with the morphology of the galaxy and evolution with redshift \citep[e.g.,][]{p06}.

The slope and normalization of the \msig\ relation are important parameters for constraining the nature of the feedback mechanism from the black hole.  
Various models and simulations generally predict $M_{bh} \propto \sigma^5$ \citep[e.g.,][]{silk98} if the feedback is in the mechanical form and $M_{bh} \propto \sigma^4$ if the feedback is dominated by momentum exchanges \citep[e.g.,][]{fabian99,k03,g04,m05}.
The normalization of the relation constrains the feedback efficiency, where $\eta \sim 0.05$ is generally required for energy feedback models and $\eta \sim 1$ for momentum feedback models.
Thus, momentum feedback models require that the majority of the black hole growth occurs in the obscured phase, which can be tested by future hard X-ray surveys.  
Energy feedback models suffer from cooling problems where the feedback energy is lost by Compton scattering with photons from AGN or other radiative cooling mechanisms \citep[e.g.,][]{k03,sn10}.
Even the 5\% feedback efficiency for energy feedback models is difficult to achieve based on current observations of AGN winds.
Observationally, there are a number of attempts to constrain the \msig\ relation (Table~1), where the slope is measured between 4--5 but with large enough errors and variety in measurements that it is unclear which feedback model drives the relation. The average slope of major studies {is $\beta=4.7$, which is inclined towards an energy-driven feedback scheme, but the samples used to measure the $M_{bh}-\sigma$ relation are still under close scrutiny.

 Considering the importance of the \msig\ relation, it is essential to thoroughly investigate the relation. Authors have previously examined the effect of the uncertainties in black hole and velocity dispersion measurements \citep[e.g.,][]{nfd06,mf01}, as well as explored selection biases in choosing samples of quasars versus nearby AGN \citep[][]{sw11}. The treatment of errors arguably can change the measured slope of the \msig\ relation, and as \citet{mf01} discuss, the limited sample size can also impact the relation. 
Recently, questions were raised whether the fitting results of the \msig\ relation are affected by selection effects and even whether the relation itself could be an artifact of selection effects. 
Studies have tended to use resolution of the sphere of influence, \ri, around the SMBH as criteria for inclusion of a galaxy in their sample as a proxy for reliable mass measurements \citep[e.g.,][]{ff05,h08}, while \citet{gultekin09} argue that no such criteria is needed. 
Restricting samples to galaxies where the sphere of influence, \ri$=G$\m$/\sigma^2$ \citep{p72}, is resolved provides a lower bound to what is observable in the \msig\ plane. 
Therefore, directly fitting these filtered samples can result in parameters different from those of an intrinsic relation \citep[e.g.,][]{gultekin09}.
Furthermore, a relation measured from a sample using this criteria might even arise simply from setting the lower limit for \msig\ data pairs. 
\citet{b10} addresses this issue by applying both the \ri\ resolution cut and the observed upper limit plus scatter in the \msig\ plane and can reproduce the \msig\ relation in simulations.  
In particular, the author uses observed \vd\ and simulated \m, evenly distributed logarithmically over $10^1$\msun\ to $10^{10}$\msun, and makes an \ri\ cut based on an Hubble Space Telescope (HST) resolution of 0\sarc1 and uses the upper scatter of the \msig\ relation itself to create an upper bound. 
This simple approach provides a physically different interpretation of the observed \msig\ relation: instead of a correlation, there is an envelope that sets an upper bound for \m\ given \vd, which may represent a more realistic model for black hole growth \citep{k10}. The lower bound therefore only arises from selection effects in this model. However, this model predicts a distribution of detection rates of black hole masses that are arguably not supported bya observations \citep{gultekin11}.

Aside from the sphere of influence selection bias, there can be another selection effect in the \vd\ distribution that is also worth exploring. 
Ideally, the relation between \m\ and \vd\ is best constrained if the individual parameters' measured distributions in a small sample follow the general distributions from a large sample.
Although it is impossible to compare to the true \m\ distribution, we find differences between the observed \vd\ distribution in the \msig\ sample and the distribution from a large sample of galaxies.
As shown in our simulation, this difference can also change the best-fit values for model parameters, especially when the underlying relation is non-linear or skewed.
In this paper, we model both the effects of the sphere of influence resolution criteria and selection effects caused by \vd\ distributions.
In addition, continuing the argument of \citet{b10}, we extend the question of whether an upper bound can be naturally produced for the observed \msig\ distribution by simulating two more general mass distributions of SMBH masses independently of the \vd\ of galaxy bulges. 
Finally, most previous studies use various direct linear regression techniques to measure the parameters of the \msig\ relation, which make it difficult to model various selection effects, and can certainly impact the final measured slope \citep{mf01}.  In this paper, we use a Bayesian Monte Carlo approach to robustly measure the parameters of the \msig\ relation including selection effects.

In Section 2, the samples used are described. Section 3 discusses methods and results, followed by conclusions in Section 4. Throughout this paper, unless otherwise noted, we assume a cosmology of $H_0$=70 km s$^{-1}$ Mpc$^{-1}$, $\Omega_M$=0.27, and $\Omega_{\Lambda}$=0.73.

\section{The Samples}
\subsection{Observed \msig\ Relation Samples}
Although there are several different compilations of measurements for the \msig\ relation \citep[e.g.,][]{ff05,gultekin09,graham11}, most of them share and/or repeat data on the same galaxies. Table~\ref{slopes} lists the major samples from the literature, with values for the parameters in the \msig\ relation when written in its standard form, log$(M_{bh} / M_{\odot})=\alpha + \beta\text{log}(\sigma /200\text{km s}^{-1})$. The variety of samples means that selecting one with which to work is not a simple process. We choose \citet[hereafter GR11]{graham11} as the main sample, with 64 galaxies. The sample contains all of the galaxies from both \citet[hereafter FF05]{ff05} and \citet[hereafter G09]{gultekin09}, with the most up-to-date values at the time. GR11 excludes four galaxies present in other samples: IC1459, for which gas and stellar dynamical models differ; NGC2748, for which dust is an issue; NGC4594, for which there is no 3-integral model; and NGC7457, where the AGN/nuclear core distinction is blurred \citep{graham11}. GR11 also includes a further 16 galaxies not previously included in the \msig\ literature. 

\begin{deluxetable}{lccccc}
\tablenum{1}
\tablewidth{0pt}
\tablecaption{Best-fit Parameters of the \msig\ Relation\label{slopes}}
\tabletypesize{\scriptsize}
\tablehead{
\multicolumn{1}{c}{} & \multicolumn{3}{c}{Literature} & \multicolumn{1}{c}{} & \multicolumn{1}{c}{$\Delta \beta$ correction} \\
\colhead{Reference} & \colhead{$\alpha$} & \colhead{$\beta$} & \colhead{$\epsilon$} & \colhead{Sample size} &  \colhead{$\beta_{intr} - \beta_{measured}$} 
}
\startdata
\citet{g00} &  $8.08\pm0.1$  &  $3.75\pm0.3$ & 0.30 &  26  &  \\
\citet{fm00}\tablenotemark{a} &  $8.14\pm1.8$  &  $4.80\pm0.54$ & \nodata & 12  & \\
\citet{fm00}\tablenotemark{b} &  $8.77\pm1.4$  &  $5.81\pm0.43$ & \nodata &  29  & \\
\citet{t02} &  $8.13\pm0.06$  &  $4.02\pm0.32$ & 0.25--0.30 &  31  & \\
\citet{ff05} &  $8.22\pm0.06$  &  $4.86\pm0.43$ & 0.34 &  25  &  \\      
\citet{h08} &  $8.18\pm0.06$  &  $4.57\pm0.37$  & 0.42 & 48  &  \\
\citet{gr08} &  $8.13\pm0.06$  &  $5.22\pm0.40$  & 0.33 & 50  &  \\
\citet{gultekin09} &  $8.12\pm0.08$  &  $4.24\pm0.41$ & 0.44 &  49(+18 upper limits) & 0.25 \\
\citet{graham11} &  $8.13\pm0.05$  &  $5.13\pm0.34$  & 0.43 & 64 & 0.76 \\
\citet{beifiori12}\tablenotemark{c} &  $8.19\pm0.07$  &  $4.17\pm0.32$  & 0.41 &  49  & 0.19 \\
\citet{mc11} & $8.29\pm0.06$ & $5.12\pm0.36$ & 0.43 & 65 & 0.50 \\
\hline  \\[-6pt]
This paper & $8.07_{-0.10}^{+0.08}$ & $5.28_{-0.55}^{+0.84}$ & $0.45_{-0.07}^{+0.08}$ & 58 & \nodata \\
\enddata
\tablenotetext{a}{$r_i$ required to be resolved.}
\tablenotetext{b}{$r_i$ not required to be resolved}
\tablenotetext{c}{Sample B of the paper, which uses secure estimates of $M_{bh}$ rather than upper limits.}
\tablecomments{Measured slope, $\beta$, and intercept, $\alpha$, for the \msig\ relation in the form log$(M_{bh} / M_{\odot})=\alpha + \beta\text{log}(\sigma /200\text{km s}^{-1})$}
\end{deluxetable}

To keep the sample volume limited to within 100 Mpc, we eliminate four galaxies from the GR11 sample, leaving a total of 60 galaxies. Two more of these galaxies are cut since they do not meet the assumed spatial resolution of our following simulations, leaving a total of 58 galaxies. Therefore, the data in our sample has an effective resolution of 0\sarc08. The velocity dispersions range from 72 km s$^{-1}$ to 335 km s$^{-1}$, while the black hole masses range from 1.1$\times10^6$\msun\ to 5.6$\times10^9$\msun. 

\subsection{Simulated Sample}
To construct a simulated sample, we start with the same method as \citet{b10}, by querying the HyperLeda\footnote{http://leda.univ-lyon1.fr/} catalogue \citep{p03} 
for galaxies with measured values of the central velocity dispersion. The search is limited to galaxies within 100 Mpc, the same limit that we place on the GR11 sample. We use two different parameters to retrieve galaxies from HyperLeda: \verb+modz+ and \verb+mod0+. The first is redshift-dependent and yields galaxies further than $\sim8$ Mpc away, while the second is redshift-independent and fills in the sample for distances down to $\sim0.7$ Mpc.  Combining these two samples gives complete coverage from very small distances all the way out to 100 Mpc. The redshift-independent distance measurements from \verb+mod0+ only captures galaxies closer than $\sim50$ Mpc, while \verb+modz+ fills in the sample with galaxies all the way up to 100 Mpc. Figure~\ref{fig:disthist} shows the distribution of distances in the sample. 
The complete base sample contains $2,870$ galaxies from $0.7-100$ Mpc, with $60\le\sigma\le400$ km s$^{-1}$, the range of interest from the observed GR11 sample. This sample is referred to as HL, and all \m\ are simulated as described in Section 3. 

\begin{figure}[H]
\figurenum{1}
\label{fig:disthist}
\begin{center}
\includegraphics[width=0.4\textwidth]{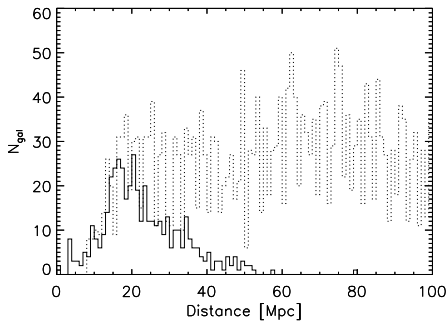}
\end{center}
\caption{The distribution of distances in the HL sample. The solid line shows the distribution for galaxies selected with mod0, and the dot-dashed line shows the distribution for galaxies selected with modz.}
\end{figure}

\section{Methods and Results}
We use Bayesian Monte-Carlo simulations to build models that can constrain and interpret the observed \msig\ distribution. Based on Bayes' theorem, given the observed distribution $\mathcal{D}$ in the \msig\ plane and the combination of selection effects $\mathcal{S}$, the probability of a hypothesis $P(\mathcal{H}|\mathcal{D}, \mathcal{S})$ is proportional to 
\begin{equation}
    P(\mathcal{H}|\mathcal{D}, \mathcal{S}) \propto P(\mathcal{D}|\mathcal{H}, \mathcal{S}) P(\mathcal{H}|\mathcal{S}),
\end{equation}
where $P(\mathcal{H}|\mathcal{S})$ is the prior probability of the model and $P(\mathcal{D}|\mathcal{H}, \mathcal{S})$ is the likelihood function. 


To construct intrinsic models $\mathcal{H}$, we use Monte-Carlo simulations assuming three different models for the black hole mass distributions: (i) an \msig\ relation characterized by three parameters, a slope $\beta$, an intercept $\alpha$, and an intrinsic scatter $\epsilon$, (ii) a power law distribution $P(M) \propto M^{-\Gamma}$, and (iii) a power law distribution with an exponential cutoff at a fixed mass of $10^9$\msun. These models can be divided into two classes of model \msig\ distributions. The first class, containing Model (i), is a simulation where the SMBH mass
for each galaxy is calculated from the HL sample σ using an intrinsic \msig\ relation, and
therefore \m\ is dependent on \vd. The second class contains two simulations, Models (ii) and
(iii), both of which generate \m\ independently of \vd. The first class of models presupposes
an intrinsic \msig\ relation, and is the focus of this paper. The second class of models is
used to explore whether selection biases alone are enough to recreate the observed \msig\
relation, and to serve as a comparison for the first class, where \m\ is a variable that is
dependent on \vd. By comparing the outcomes of the two classes, we can comment on which
outcome is more likely, an intrinsic relation, or a relation that arises only because of selection effects. In all cases, we use the HL sample with 2,870 galaxies for our \vd, and simulate
SMBH masses for each galaxy. For each distribution, we focus on the mass range of $10^6$\msun\ to $10^{10}$\msun.

Once the intrinsic models are constructed, we apply selection effects, or selection functions $\mathcal{S}$. The first selection function is the familiar sphere of influence cut ($\mathcal{S}_{r_i}$), which selects galaxies where the ratio of the sphere of influence to the resolution is $\ge1$. We use a resolution of 0\sarc08, which is close to the resolution of 0\sarc1 that \citet{gultekin09} and \citet{b10} use. This provides a slightly looser constraint on the lower bound, but is still enough to identify implications from this selection effect. 
The second selection function applied, $\mathcal{S}_{\sigma}$, is dependent on the velocity dispersion distribution.
Ideally, the velocity dispersion distribution of the observed sample should follow the general distribution for all galaxies to reduce the selection effect, especially if the underlying relation is non-linear or distorted.
Because the general \msig\ relation is measured for a combination of different type of galaxies, we use the velocity dispersion distribution from HL, a large sample of galaxies, as a proxy for our comparison.  
Figure~\ref{hlhist} shows the velocity dispersion distributions in the HL and GR11 samples, as well as the velocity dispersion function for early-type galaxies from \citet{sheth03}. 
We find that in the observed sample (GR11) there are relatively more black mass measurements at the high velocity dispersion regime, and less in the low velocity dispersion regime. 
A Kolmogorov-Smirnov (K-S) test comparing the GR11 and HL samples returns a p-value of 0.008, indicating that the samples are consistent only within 3$\sigma$ of each other. 
Comparing the HL sample and the velocity dispersion function for early-type galaxies, the velocity dispersion for early-type galaxies is narrower than the HL sample. 
Therefore, the velocity dispersion distribution in the observed sample for measuring the \msig\ relation does not follow that of a large sample of galaxies, and we include this selection effect in our simulation to test if this difference can cause significant changes in our final parameter estimations for intrinsic models.
To match the distribution for the GR11 sample (applying the $\mathcal{S}_{\sigma}$ selection), we divide the \vd -space into five equal bins, and then $\mathcal{S}_{\sigma}$ cuts the number of galaxies in each bin from the model $\mathcal{H}$, such that the number ratios are consistent with those of the GR11 sample. 
The third selection function is a combination of both the sphere of influence cut and velocity dispersion sample bias, and is denoted as $\mathcal{S}_{\sigma,r_i}$.

\begin{figure}[H]
\figurenum{2}
\begin{center}
\includegraphics[width=3in,clip]{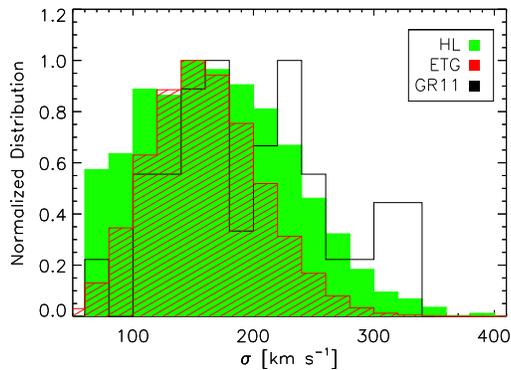}
\end{center}
\caption[Velocity Dispersion Distributions]{Histograms showing the normalized distributions of velocity dispersion for the HyperLeda (solid shaded green area), early type galaxy function from \citet[][hashed red area]{sheth03}  and \citet[][solid black line]{graham11} samples. The HL and GR11 samples are only consistent within the 3$\sigma$ limit.}
\label{hlhist}
\end{figure}

After simulating a model and applying selection effects, we obtain a large number of scattered data pairs, and we compare the observed \msig\ data pairs with the simulated data pairs to calculate the likelihood function $P(\mathcal{D}|\mathcal{H}, \mathcal{S})$.
First, we use the two-dimensional K--S test with and without rotations to calculate the likelihood function.  However, we find the K--S test is too general for this specific case, where the observed data fall in a linear relation, and the resulting constraints from our simulations are not tight.
Since the sphere of influence selection causes the resultant simulated data points to always look linear, we use a linear model parameterized by a slope, an intercept, and an intrinsic scatter ($\beta', \alpha', \epsilon'$) to model the simulated data.  Essentially, we assume $P(\mathcal{D}|\mathcal{H}, \mathcal{S})$ = $P(\mathcal{D}|\beta', \alpha', \epsilon')$ and evaluate $P(\mathcal{D}|\beta', \alpha', \epsilon')$ using the standard $\chi^2$ likelihood including intrinsic scatter.
Comparing the two methods, we find that the final expected values from posterior probability distributions are quite consistent; however, the uncertainties using the K--S method are always larger.  Therefore, we subsequently only discuss results from the second approach.

Since there is no prior knowledge on the model parameters given the selection effects, the prior $P(\mathcal{H}|\mathcal{S})$ is essentially proportional to the distributions of the model parameters, e.g., $\beta$ in the \msig\ relation, where we assume uniform distributions for the model parameters. 
To explore the three dimensional parameter space with the \msig\ simulations, we use ranges of $7.8\le\alpha\le9$ for the intercept, $1\le\beta\le10$ for the slope, and $0.1\le\epsilon\le1$ for the scatter. All of these ranges use a step size of 0.01, and  cover the expected area of interest.  For power law and exponential cutoff distributions, we assume a power law slope range of $1.0 < \Gamma < 1.6$.  For model parameters outside of these ranges, the likelihood function is close to zero and will not contribute to the final parameter estimation.
Finally, we measure the posterior probability distributions, $P(\mathcal{H}|\mathcal{D}, \mathcal{S})$, using Equation~1, and estimate the expected model parameters as $\bar{X}=\sum\limits_i P_iX_i / \sum\limits_i P_i$ and $1\sigma$ uncertainties from $P(\mathcal{H}|\mathcal{D}, \mathcal{S})$.

\subsection{Model (i): Assuming an Intrinsic \msig\ Relation}
We probe the entire three dimensional parameter space by letting $\alpha$, $\beta$, and $\epsilon$ vary, calculate the $\chi^2$ likelihood of each individual combination of parameters, and construct the posterior probability distributions.  We compute the expected value for the intercept, slope, and scatter  $\bar{X}=\sum\limits_i P_iX_i / \sum\limits_i P_i$ from the 3D posterior probability distribution. 
We marginalize the other two parameters for a particular parameter, $X$, by summing all the probabilities for a specific value of $X$ (Figure~\ref{fig:exp}). 
Figure~\ref{fig:exp} shows the marginalized posterior probability distributions for each parameter for each application of selection effects, and where the expected values for each parameter of the simulation are marked. Integrating under each curve gives the 68\% limits on the expected values. 

\begin{figure}[H]
\figurenum{3}
\label{fig:exp}
\begin{center}
\subfigure{\includegraphics[width=0.3\textwidth ,clip]{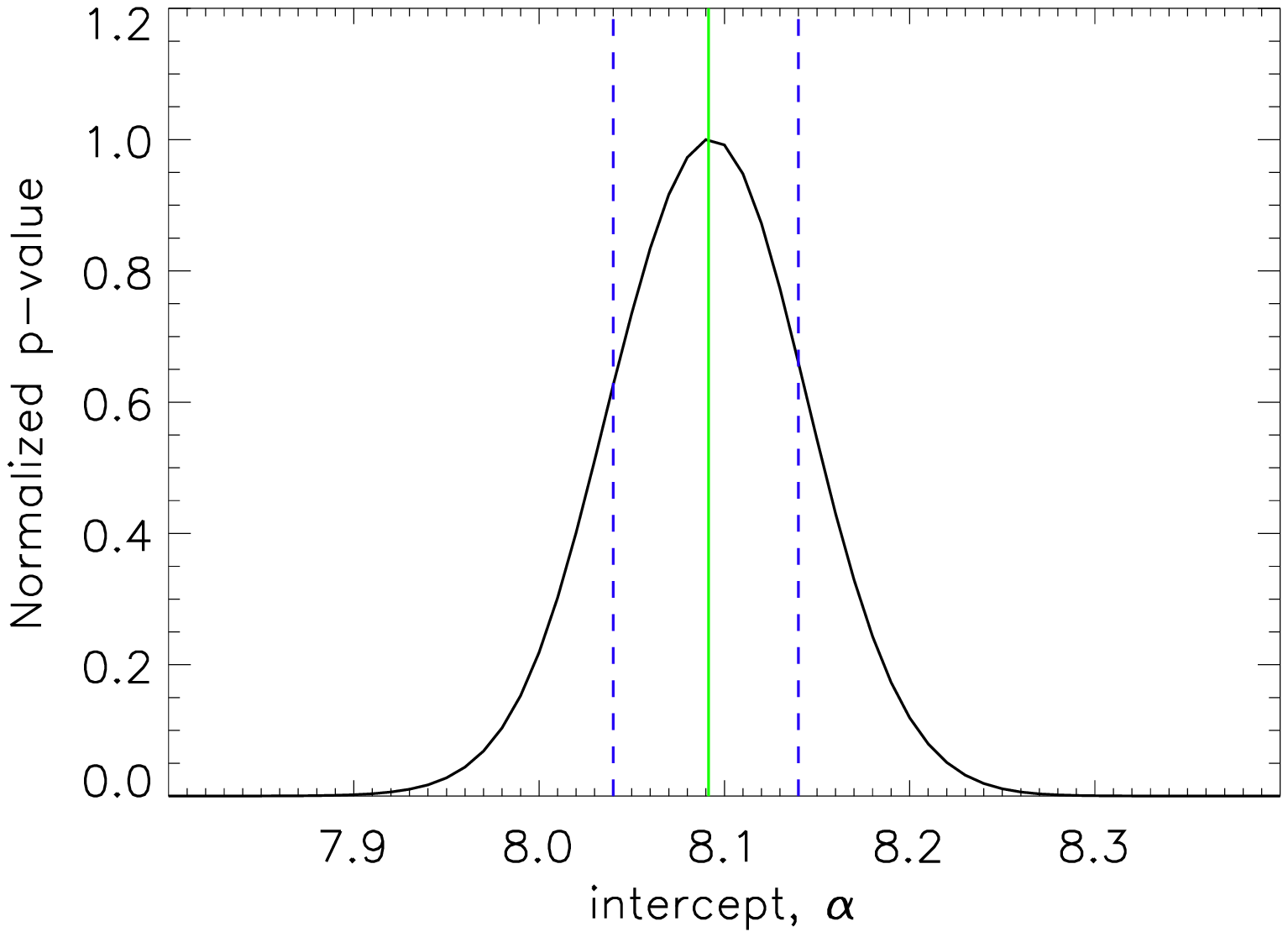}}
\subfigure{\includegraphics[width=0.3\textwidth ,clip]{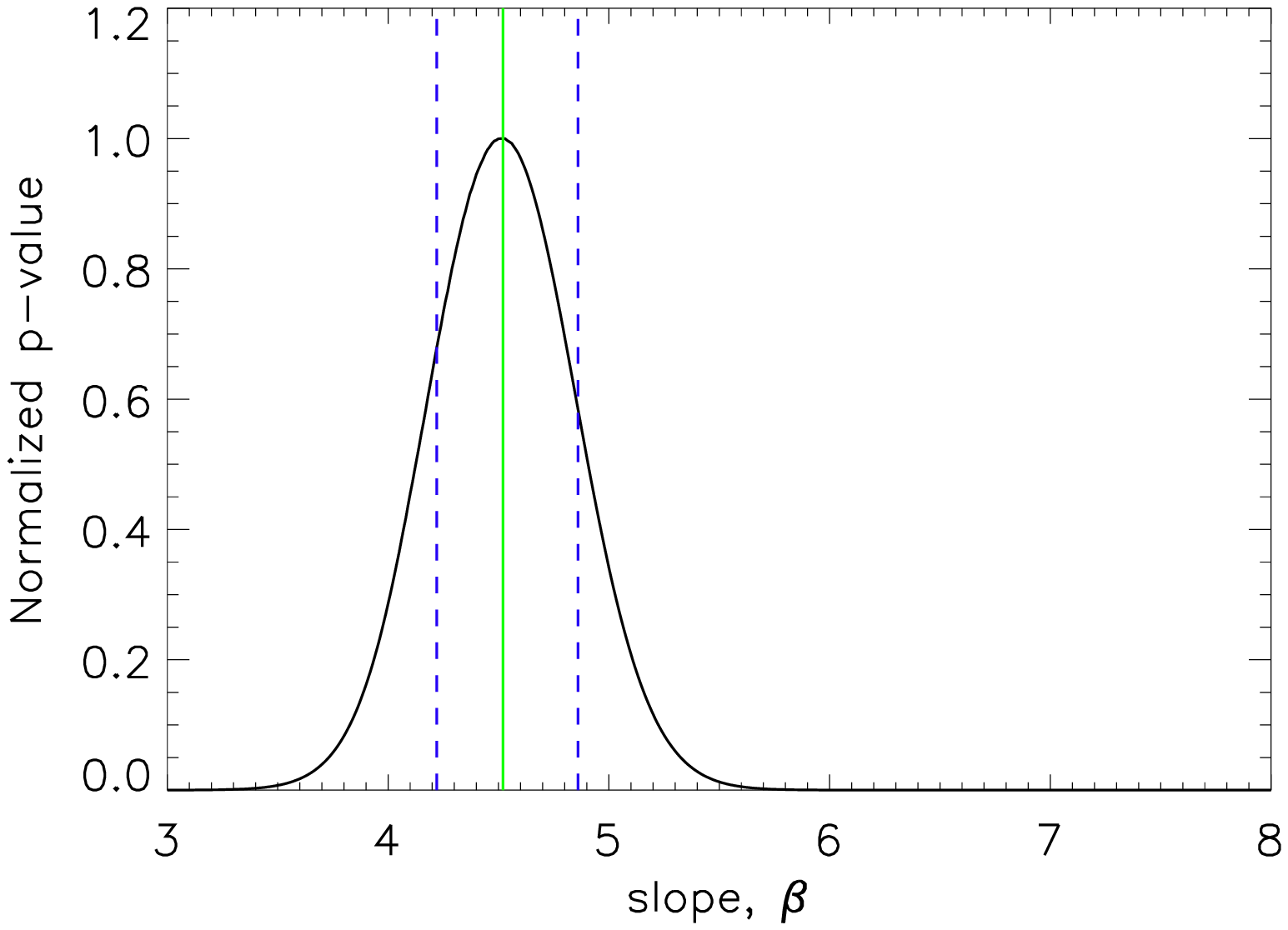}}
\subfigure{\includegraphics[width=0.3\textwidth ,clip]{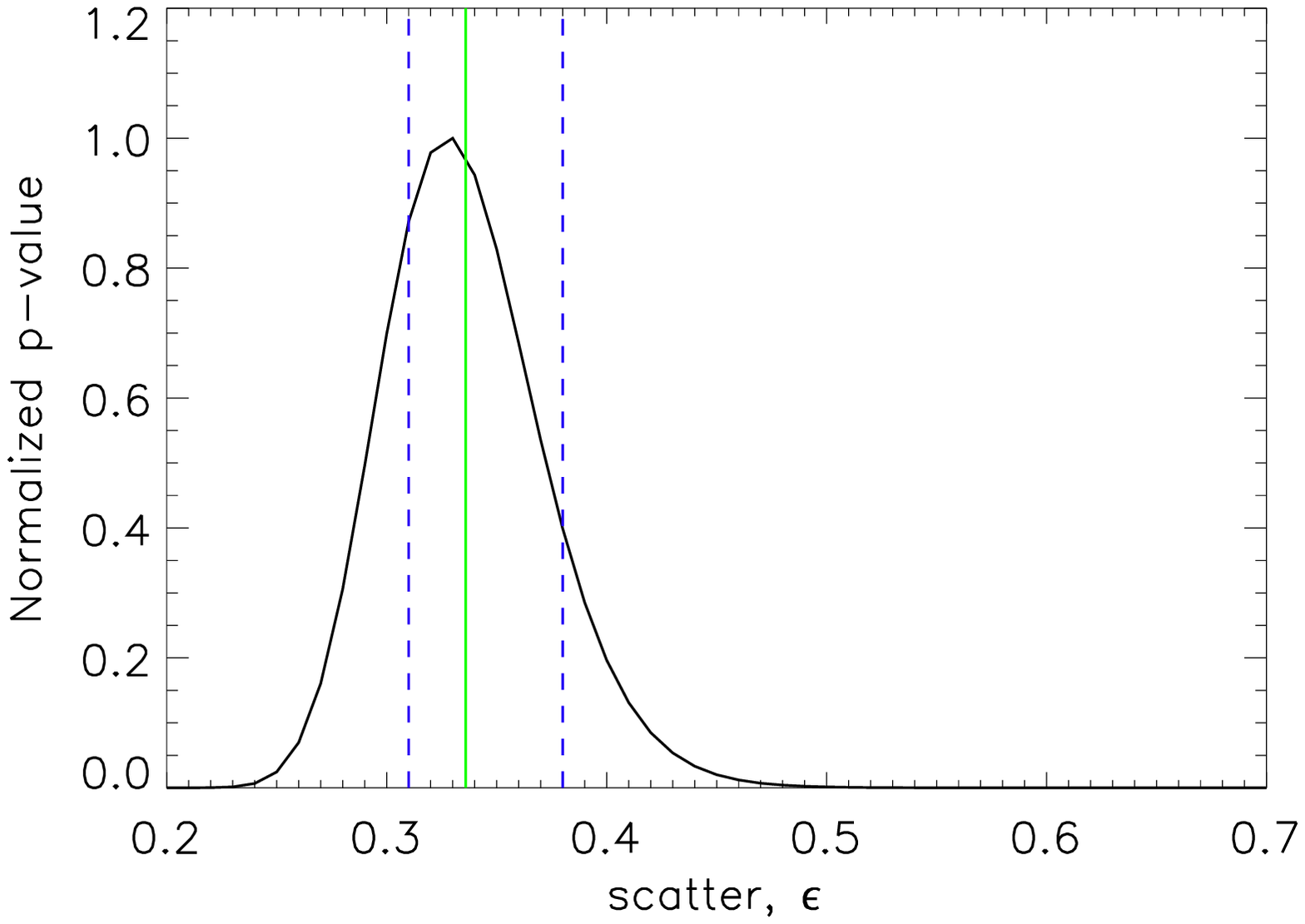}}
\subfigure{\includegraphics[width=0.3\textwidth ,clip]{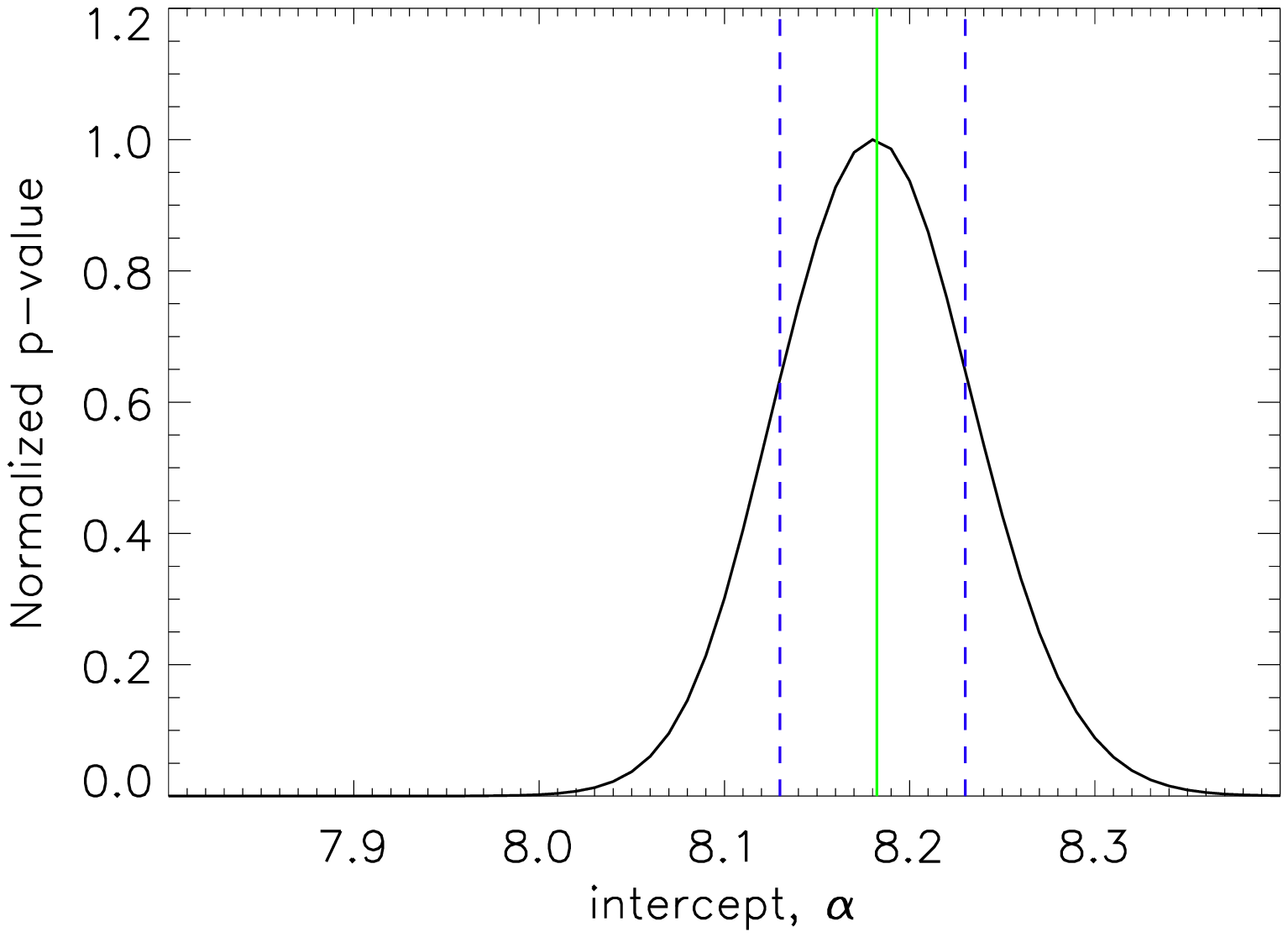}}
\subfigure{\includegraphics[width=0.3\textwidth ,clip]{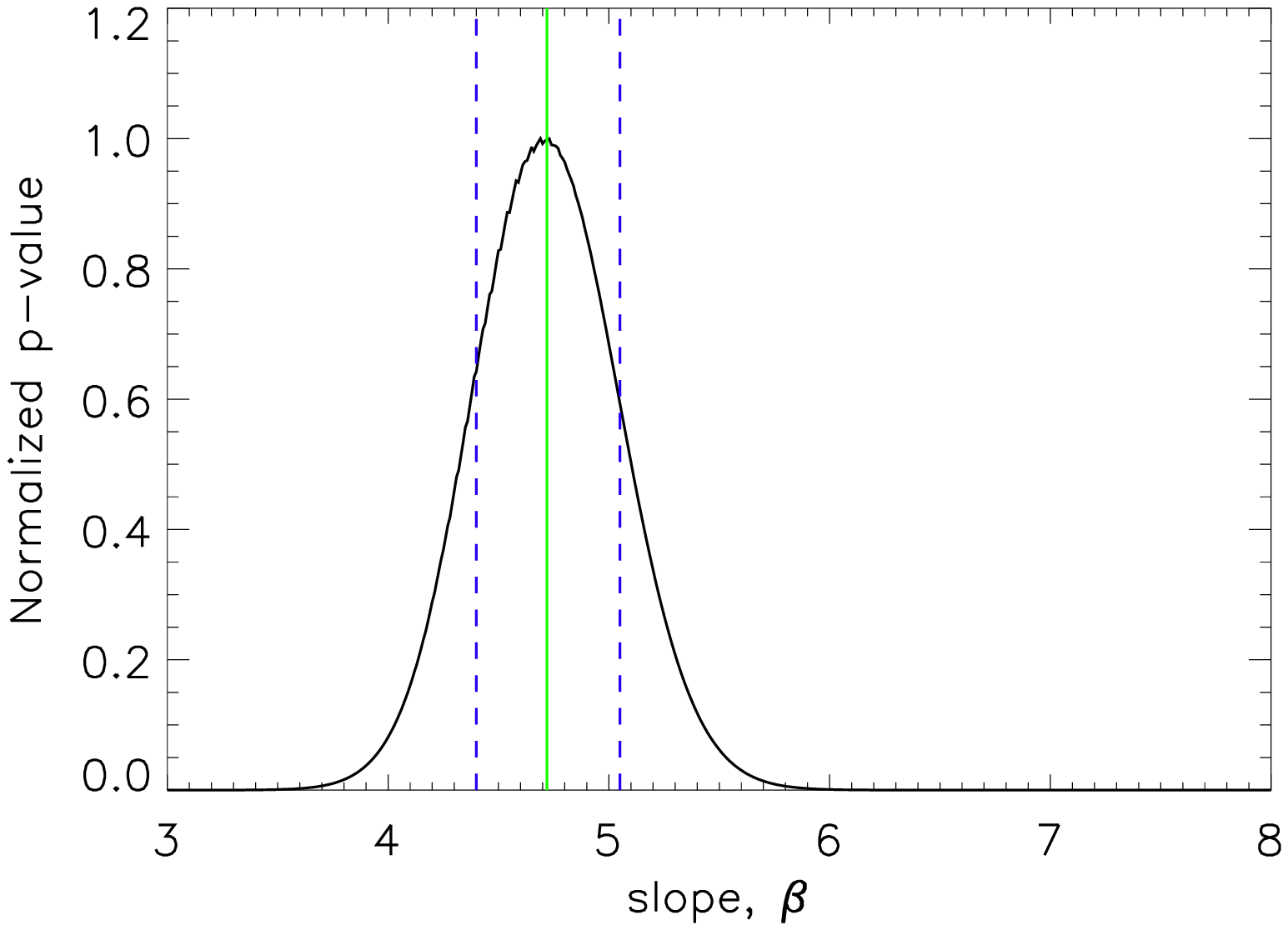}}
\subfigure{\includegraphics[width=0.3\textwidth ,clip]{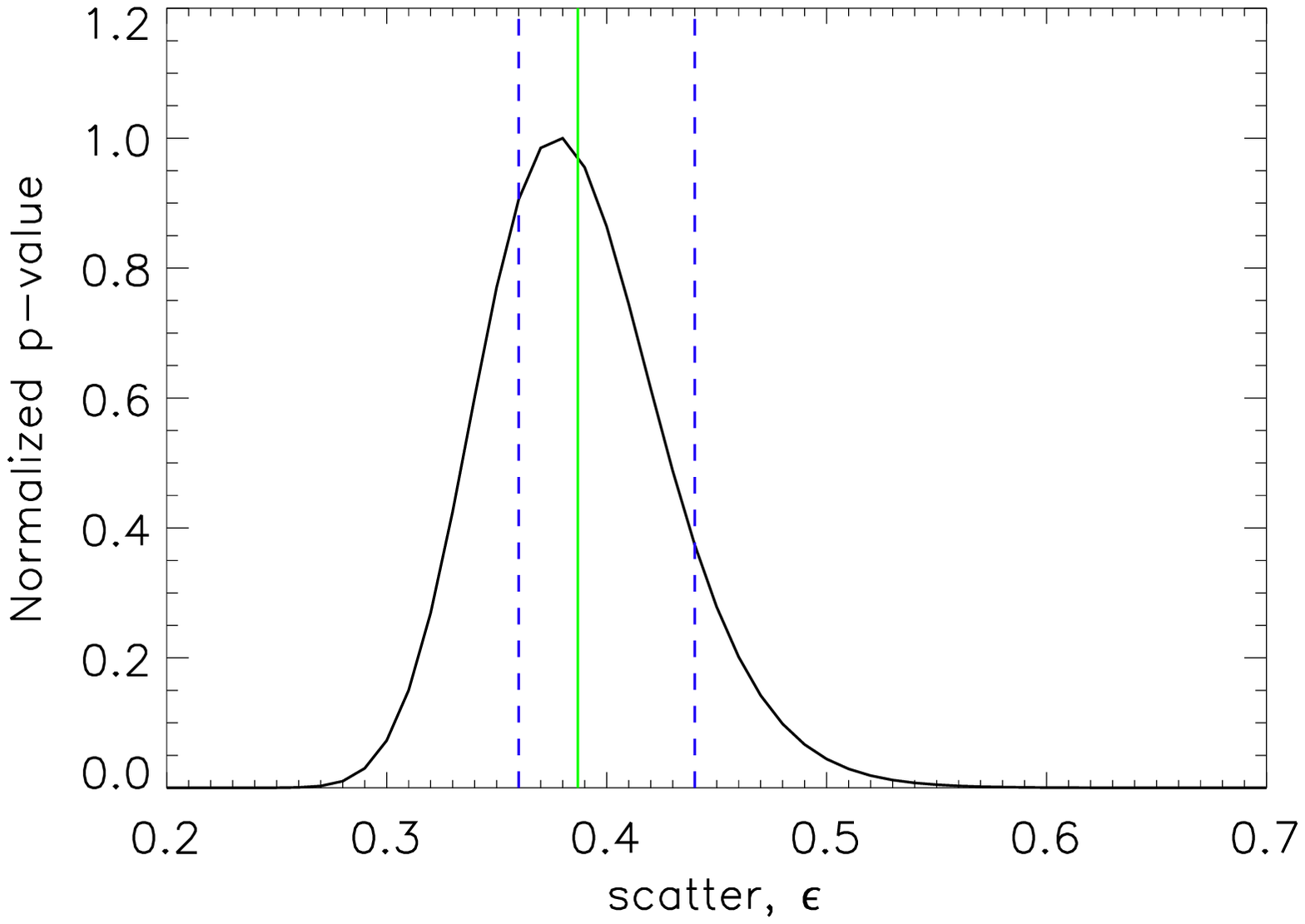}}
\subfigure{\includegraphics[width=0.3\textwidth ,clip]{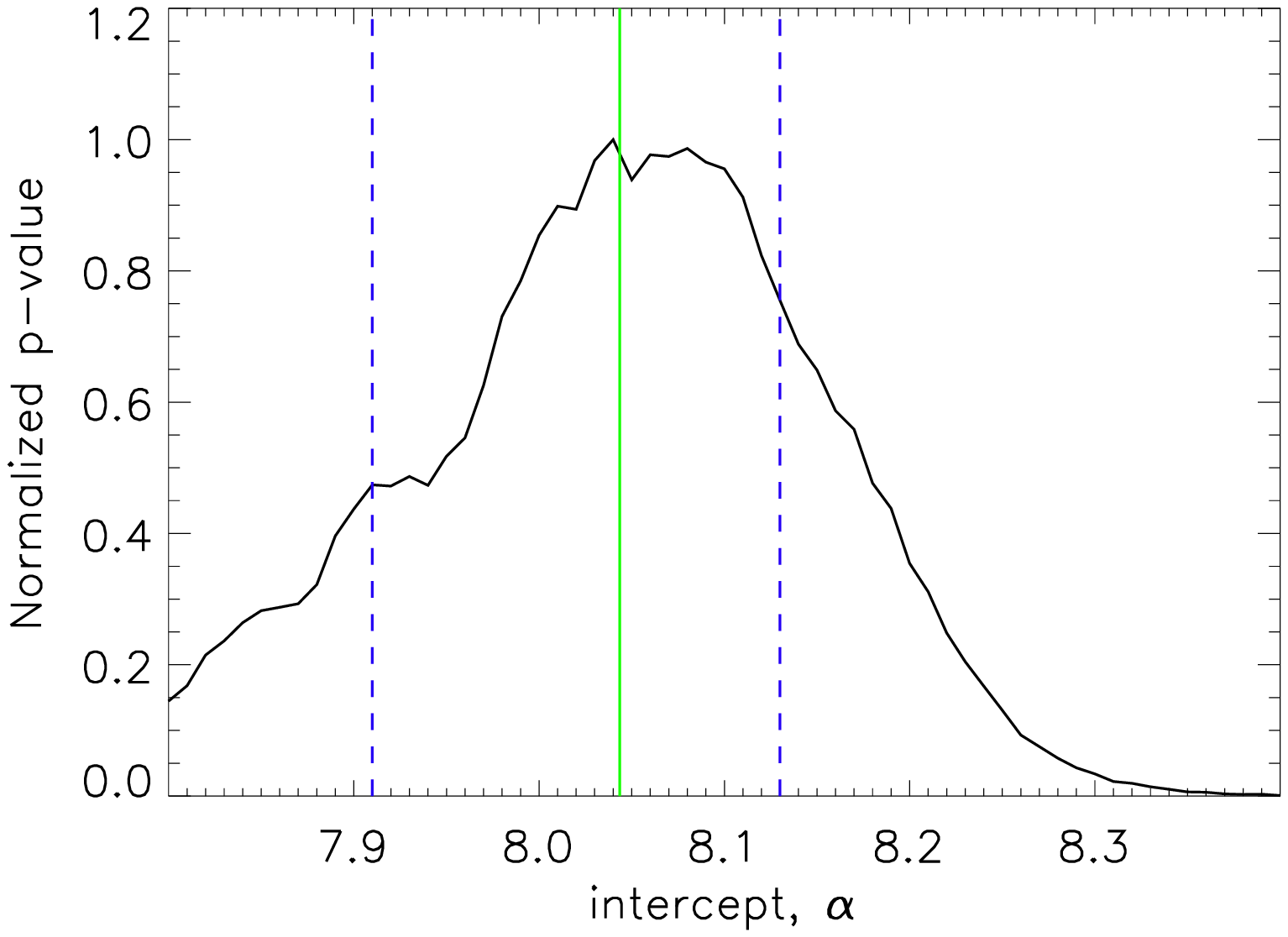}}
\subfigure{\includegraphics[width=0.3\textwidth ,clip]{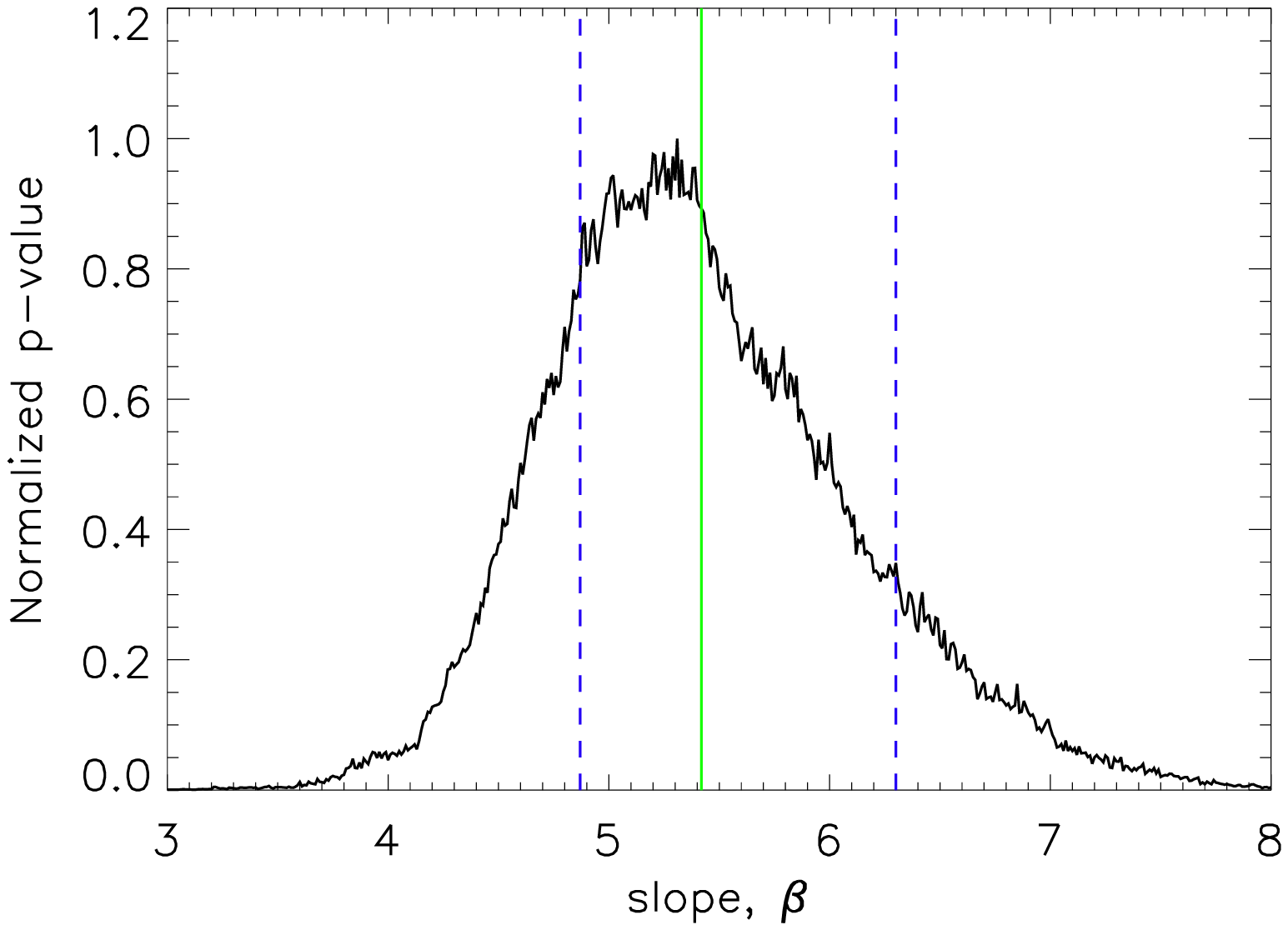}}
\subfigure{\includegraphics[width=0.3\textwidth ,clip]{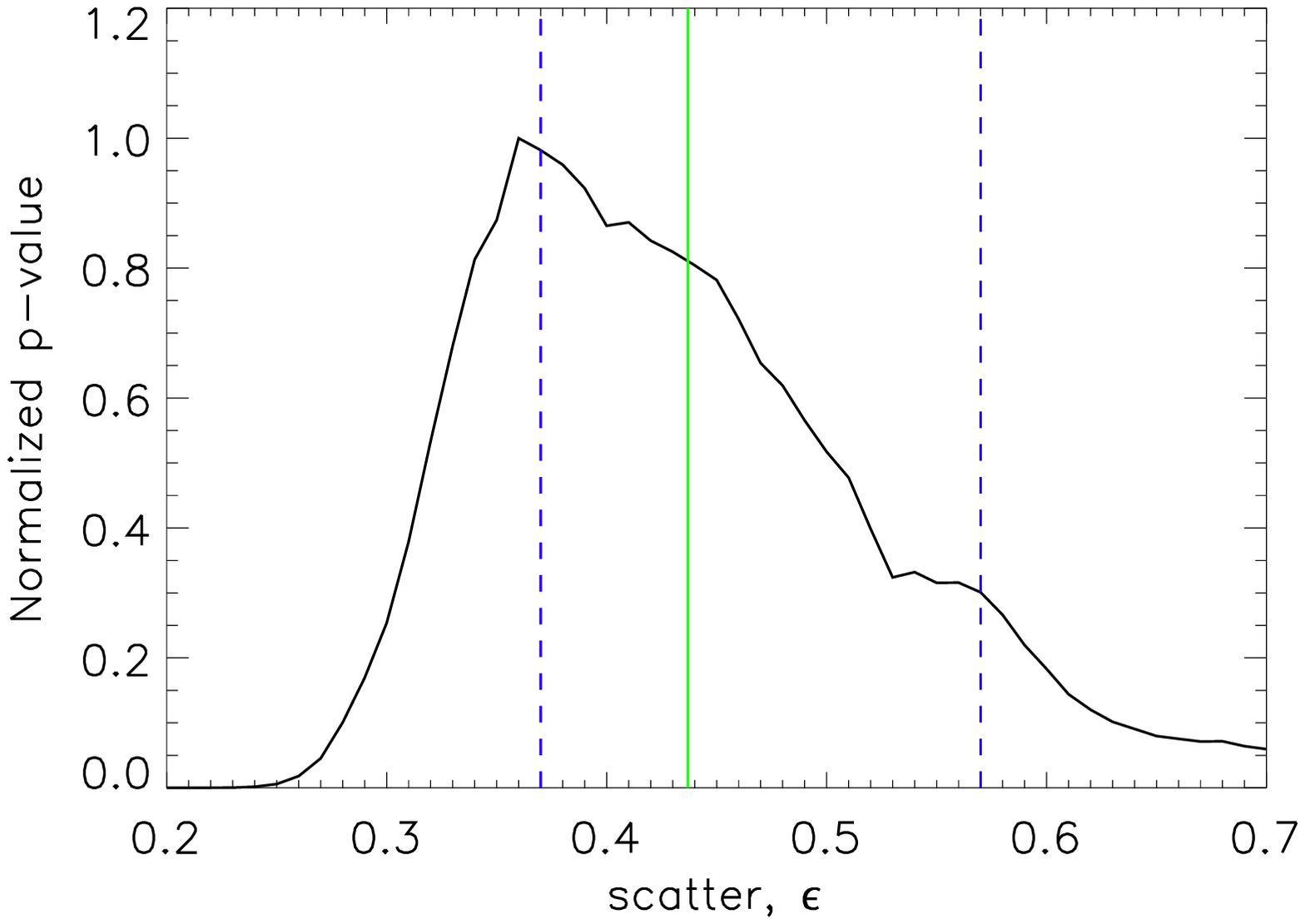}}
\subfigure{\includegraphics[width=0.3\textwidth ,clip]{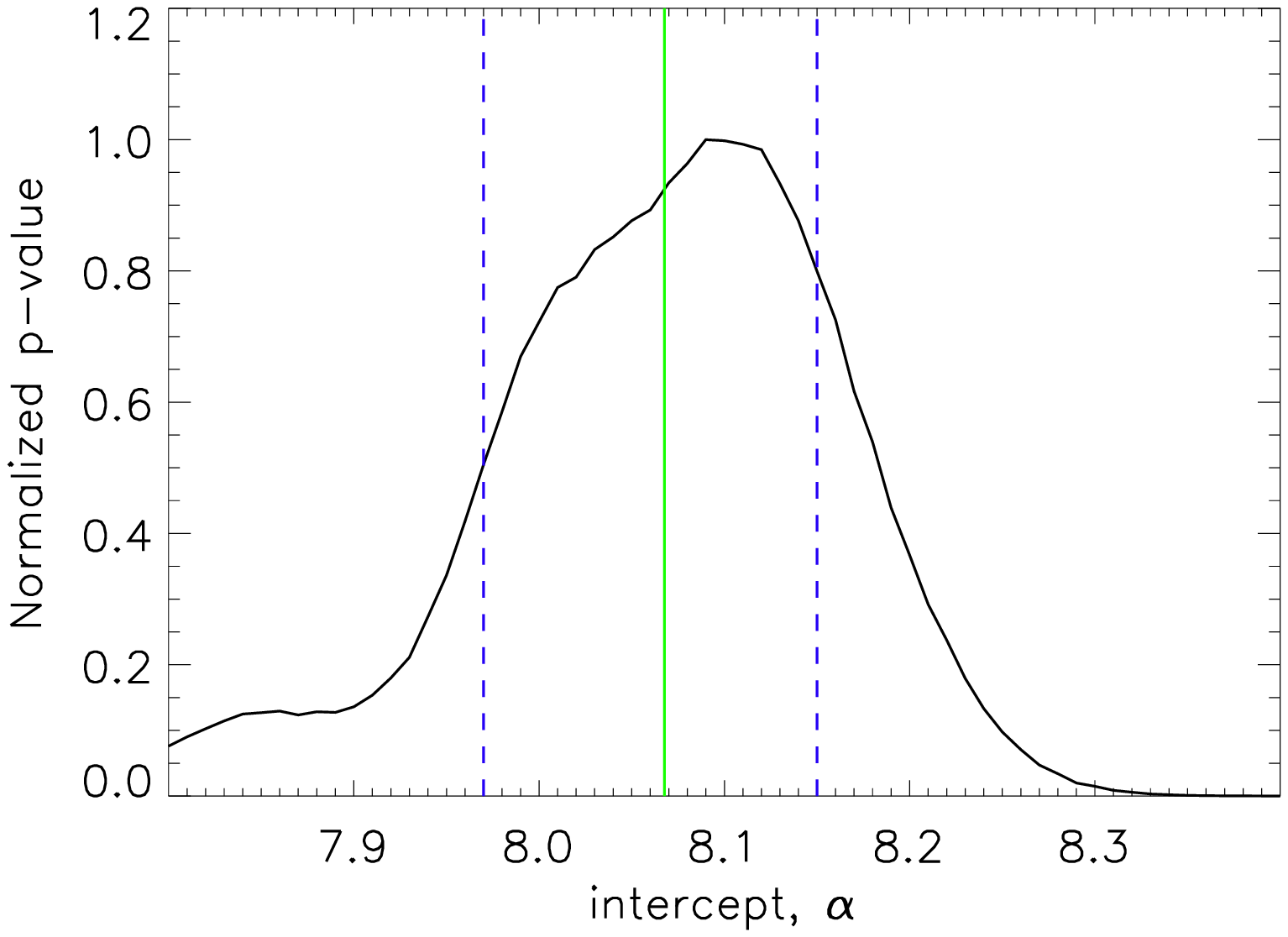}}
\subfigure{\includegraphics[width=0.3\textwidth ,clip]{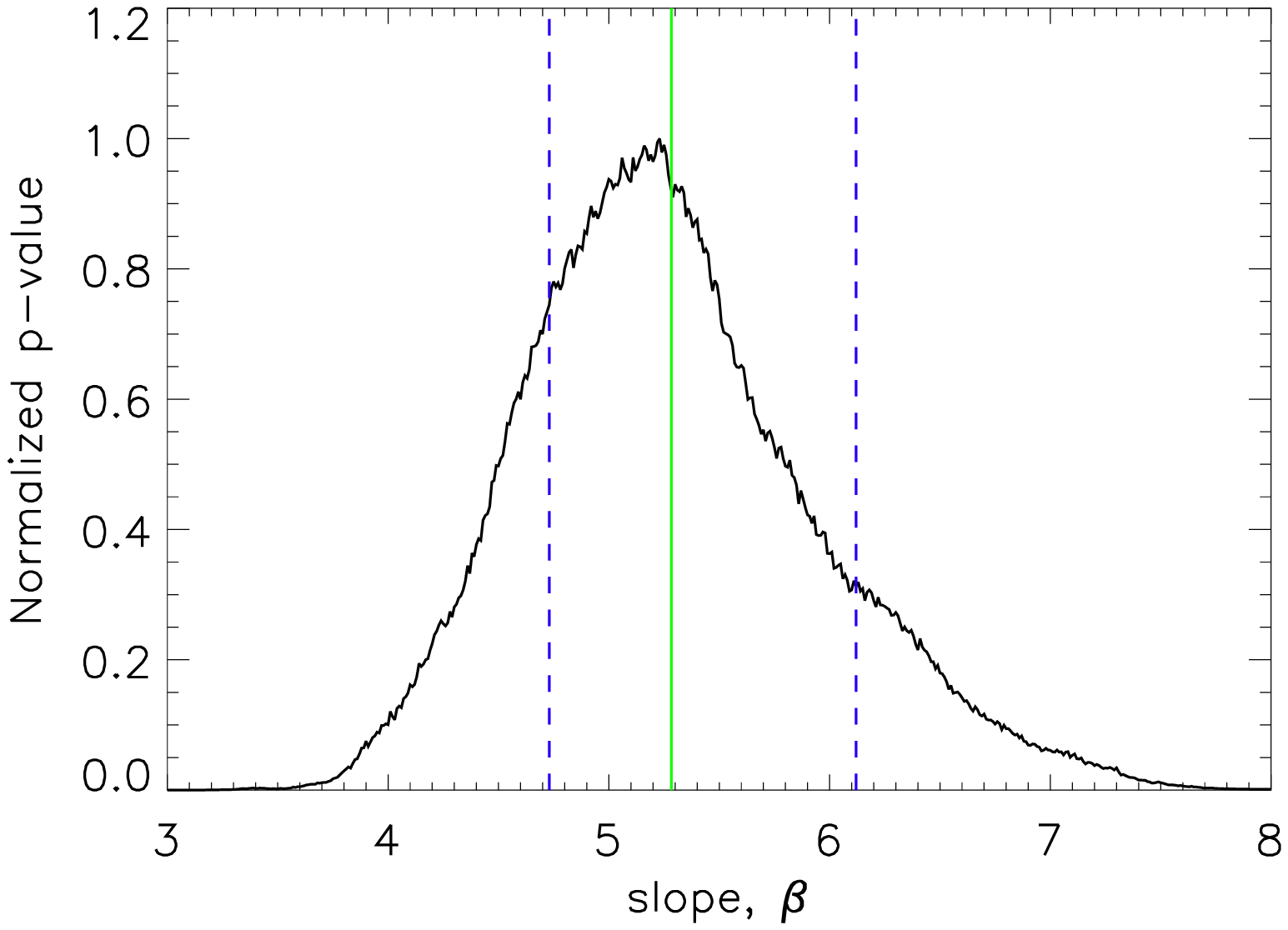}}
\subfigure{\includegraphics[width=0.3\textwidth ,clip]{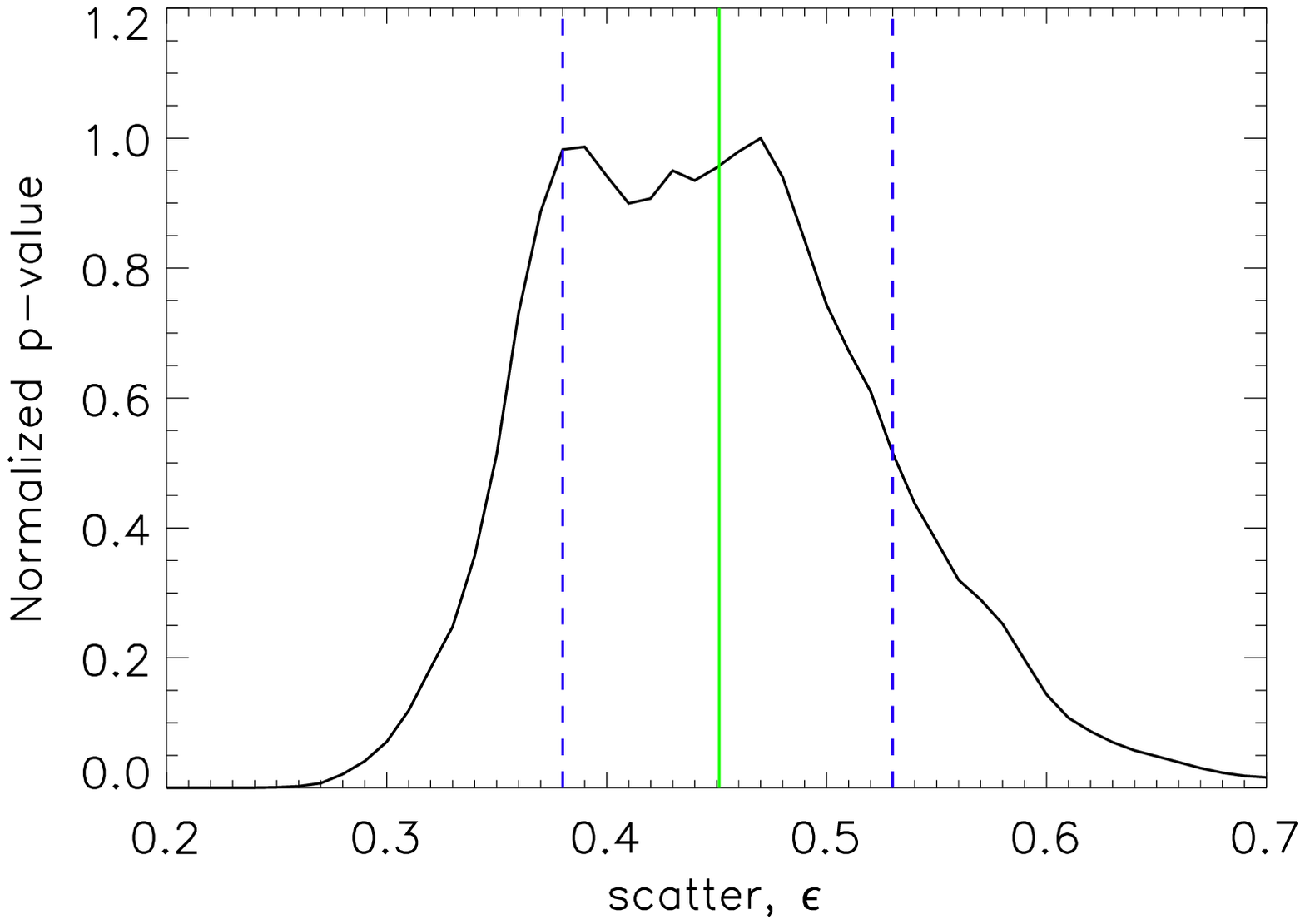}}
\end{center}
\caption[Expected Values for Intrinsic \msig\ Relation]{Marginalized posterior probability distributions for the three parameters that describe the \msig\ relation, $\alpha$, $\beta$, and $\epsilon$. The left column shows the expected values for the intercept $\alpha$, the middle column for the slope $\beta$, and the right column for the scatter $\epsilon$. The top row is the unfiltered simulation, the second row shows the simulation with the velocity dispersion selection bias ($\mathcal{S}_{\sigma}$), the third row the sphere of influence selection effect ($\mathcal{S}_{r_i}$), and the bottom row with both selection effects ($\mathcal{S}_{\sigma,r_i}$) applied. The green, solid vertical line is the expected value, and the 68\% limits are depicted by the blue, dashed vertical lines. The summed probabilities have been normalized to have a peak value of unity.}
\end{figure}

\begin{figure}
\figurenum{4}
\label{fig:expbeta}
\begin{center}
\subfigure{\includegraphics[width=0.3\textwidth ,clip]{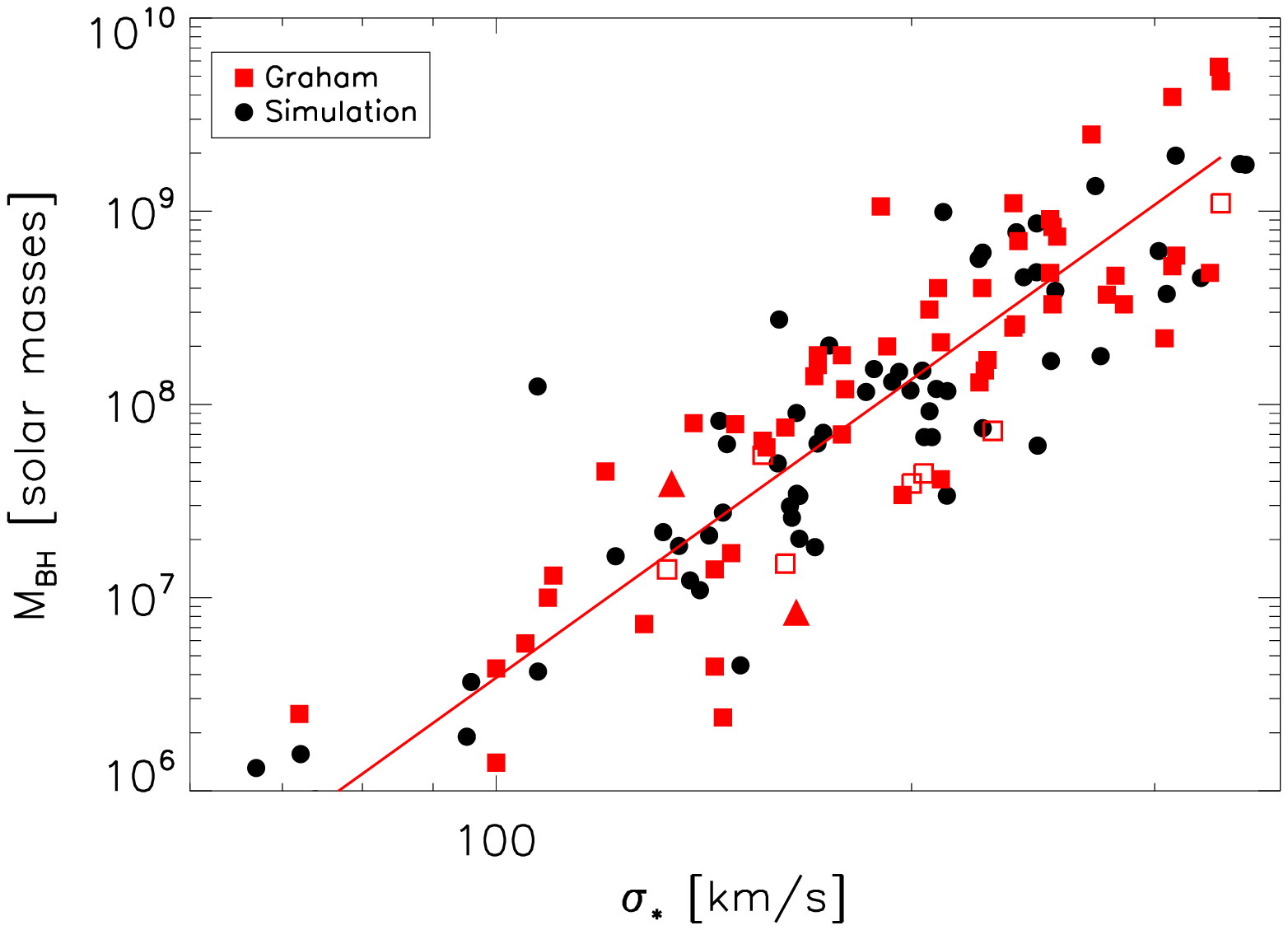}}
\subfigure{\includegraphics[width=0.3\textwidth ,clip]{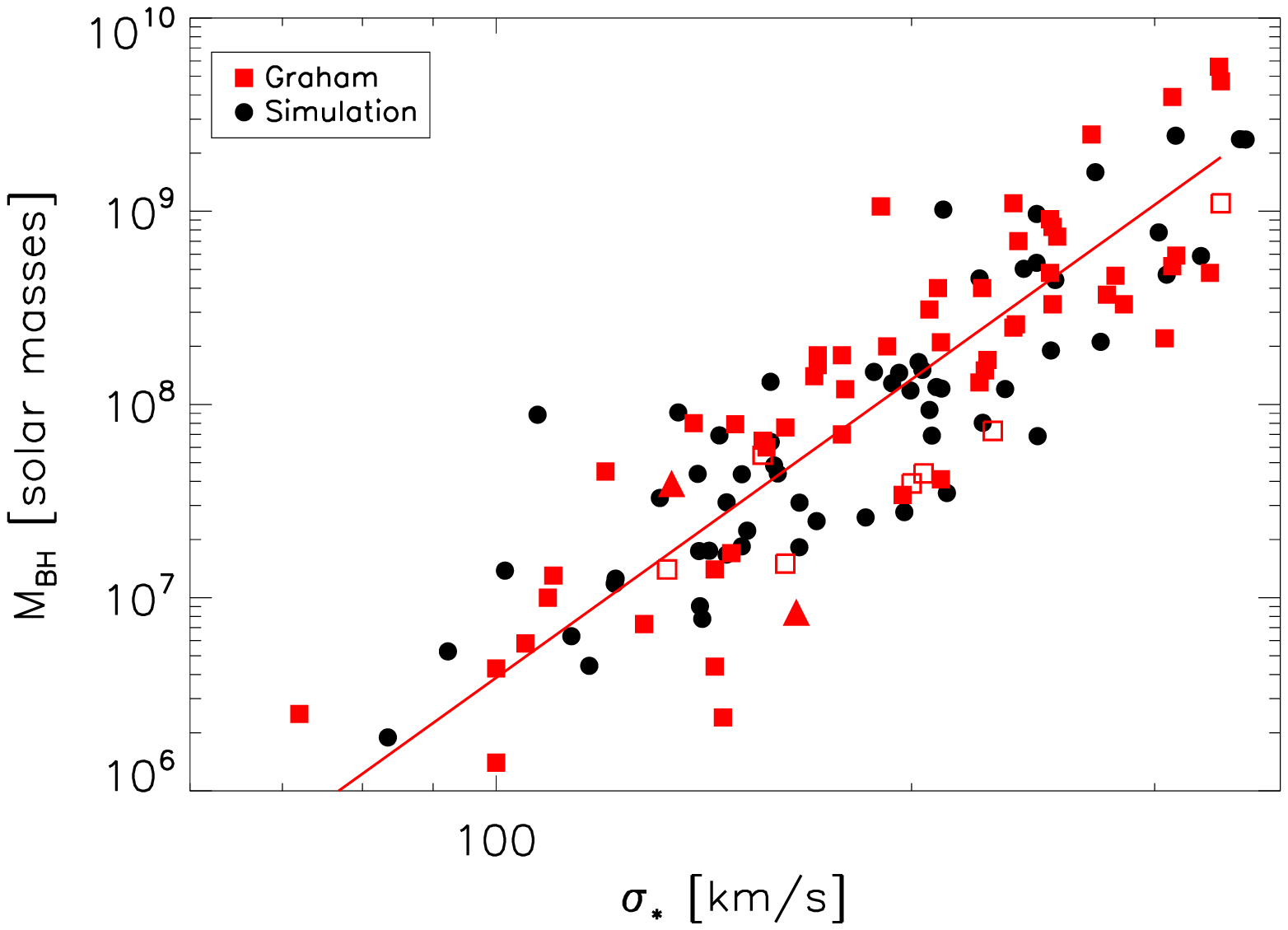}}
\subfigure{\includegraphics[width=0.3\textwidth ,clip]{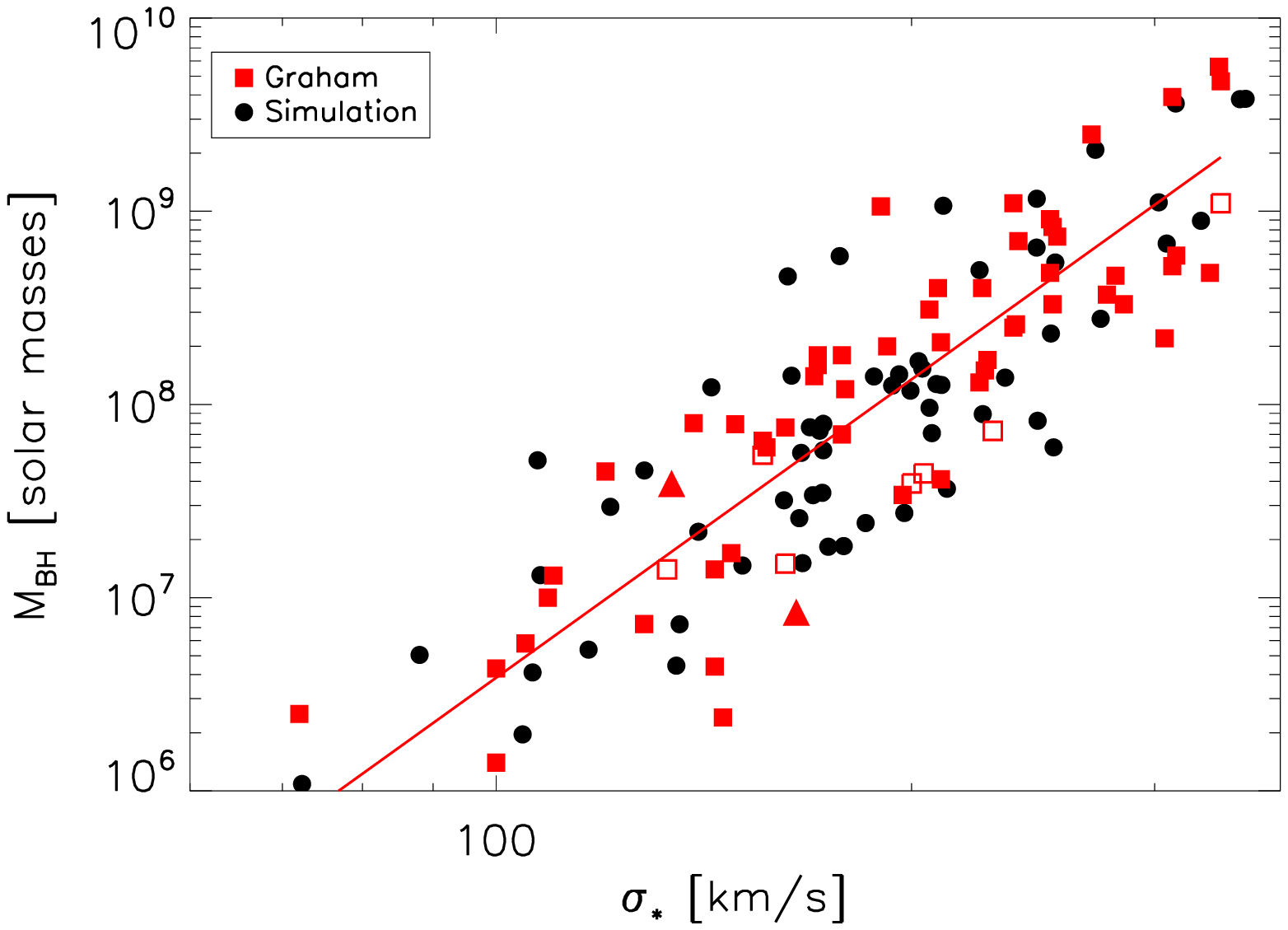}}
\end{center}
\caption[Simulation Results]{Simulation results plotted with both selection effects applied. The center panel shows the simulation for the expected value of $\beta = 5.28$, while the left and right show cases of simulations ($\beta = 4.74$ and $\beta = 6.15$) corresponding to the 68\% limits of the expected value. The black circles are from our simulations, while the red squares correspond to points in the GR11 sample. Information on the method used (triangles for masers, squares for gas or stellar kinematics) and resolution of the sphere of influence (filled symbols for resolved, open for unresolved) comes from G09 and FF05.}
\end{figure}

The expected values for the parameters are listed in Table~\ref{tab:exp}. It is important to note that these are the expected values for the \emph{intrinsic} parameters from posterior probability distributions, i.e., what is used to generate the simulation that populates the \msig\ space. When selection effects are applied, the measured parameters from a direct fitting method without modeling the selection effects can be different from intrinsic parameters.  For example, for a given intrinsic slope, the directly measured slopes are lower than the intrinsic slope for $\beta > 4$. 
This can be seen in Figure~\ref{fig:inout}, where the directly measured value $\beta'$ is plotted against the intrinsic slope $\beta$ that generated the intrinsic model. 
The selection function including both selection effects has the largest influence on the intrinsic model. The selection effects, therefore, cause a measurement of slope that is significantly lower than the intrinsic slope for $\beta>4$. For example, \citet{ff05} measure the slope to be $4.86\pm0.43$ when applying the sphere of influence resolution criteria, which our simulations show would have an intrinsic slope of $\beta\sim6$. This is close to the slope of $5.81\pm0.43$ that \citet{ff05} measure without the sphere of influence selection effect, supporting the results of our simulations.

\begin{figure}
\figurenum{5}
\label{fig:inout}
\begin{center}
\includegraphics[width=0.5\textwidth]{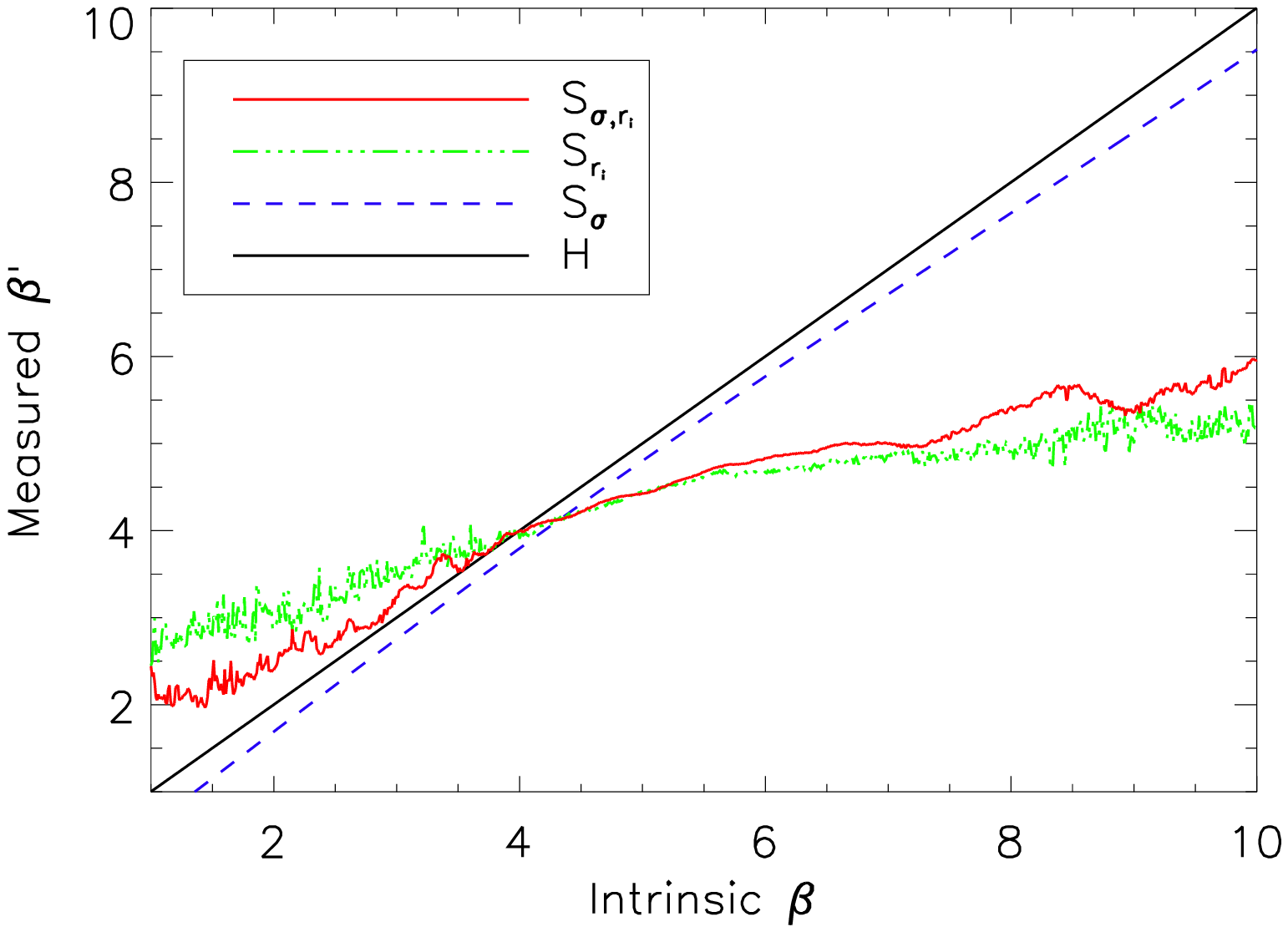}
\end{center}
\caption[Measured vs. Expected Slope]{The directly measured slope, $\beta'$, vs. the intrinsic $\beta$ of the simulation. The curve with no selection effects has a slope of unity, as expected. The selection effects cause the directly measured slope to be different than that of the intrinsic model.}
\end{figure}

\begin{deluxetable}{lccc}
\tablenum{2}
\tablewidth{0pt}
\tablecaption{\msig\ Expected Values\label{tab:exp}}
\tabletypesize{\scriptsize}
\tablehead{\colhead{} & \colhead{$\bar\alpha$} & \colhead{$\bar\beta$} & \colhead{$\bar\epsilon$}  }
\startdata
$\mathcal{H}$ & $8.09_{-0.05}^{+0.05}$ & $4.52_{-0.30}^{+0.34}$ & $0.34_{-0.03}^{+0.04}$ \\[2pt]
$\mathcal{H},\mathcal{S}_{r_i}$ & $8.18_{-0.05}^{+0.04}$ & $4.72_{-0.32}^{+0.33}$ & $0.39_{-0.03}^{+0.05}$ \\[2pt]
$\mathcal{H},\mathcal{S}_{\sigma}$ & $8.04_{-0.12}^{+0.09}$ & $5.42_{-0.55}^{+0.88}$ & $0.44_{-0.07}^{+0.13}$  \\[2pt]
$\mathcal{H},\mathcal{S}_{\sigma, r_i}$ & $8.07_{-0.10}^{+0.08}$ & $5.28_{-0.55}^{+0.84}$ & $0.45_{-0.07}^{+0.08}$ \\[2pt]
\enddata
\tablecomments{Expected values from the 3D posterior probability space of the \msig\ relation using Bayesian Monte-Carlo simulations.}
\end{deluxetable}

To illustrate the correlation between the model parameters, for each 2D parameter pair we marginalize the third dimension and plot the confidence level contours in Figure~\ref{fig:con}. The confidence levels are found by integrating along isometric contours under the 2D surface. For example, the top left panel of Figure~\ref{fig:con} shows $\alpha$ vs.\ $\beta$ with the third dimension, $\epsilon$, marginalized, for the base simulation without any selection functions applied. The expected values, $\bar\alpha,\bar\beta$ are marked by a star and the peak position of the distribution is marked with a circle. We then integrate under the surface from the peak position until reaching 68\% of the total volume under the surface, providing the $1\sigma$ contour. The contours become larger as selection effects are applied, increasing the amount of error that is introduced into the measurement of the \msig\ relation.

\begin{figure}
\figurenum{6}
\label{fig:con}
\begin{center}
\subfigure{\includegraphics[width=0.3\textwidth ,clip]{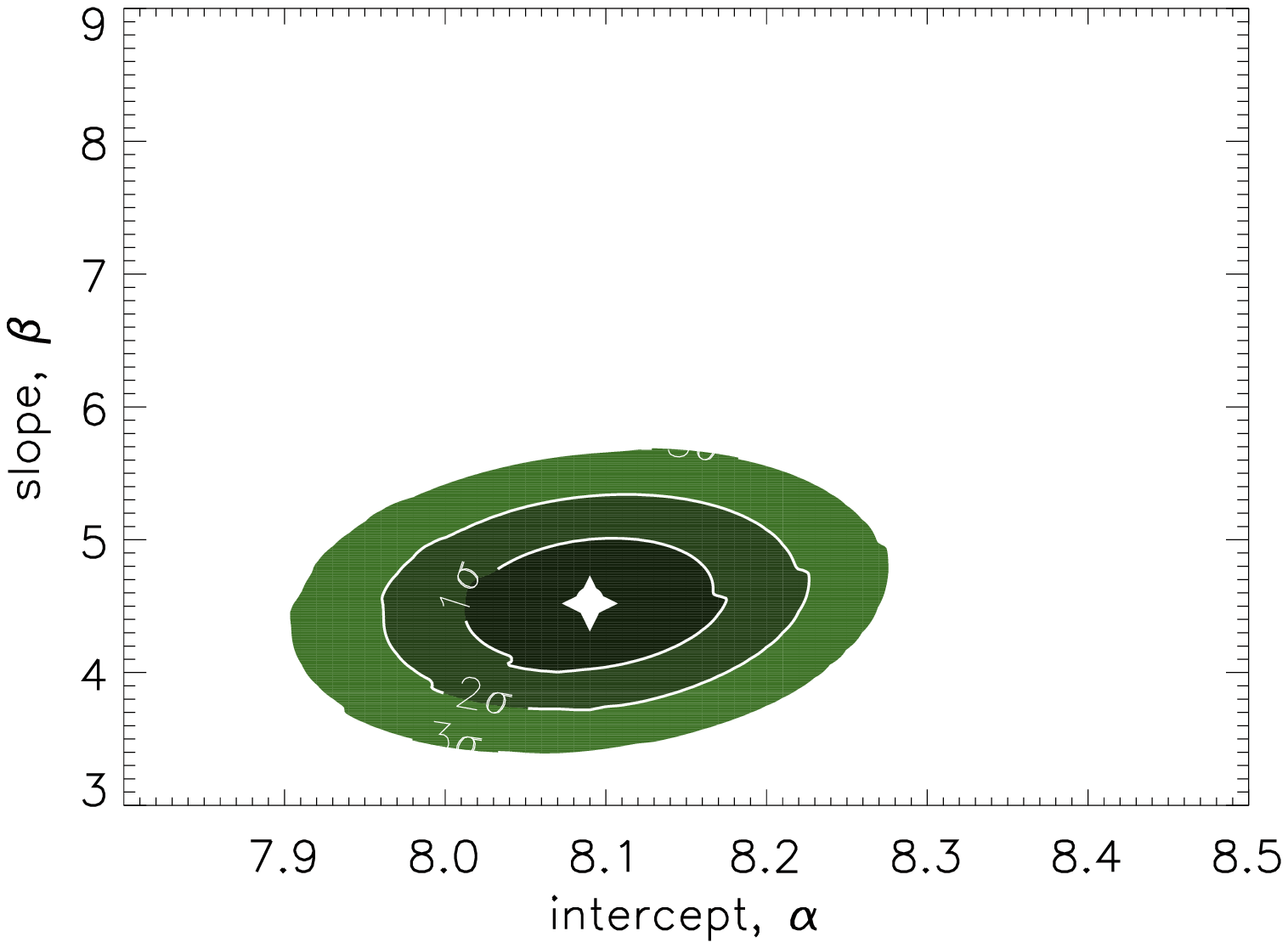}}
\subfigure{\includegraphics[width=0.3\textwidth ,clip]{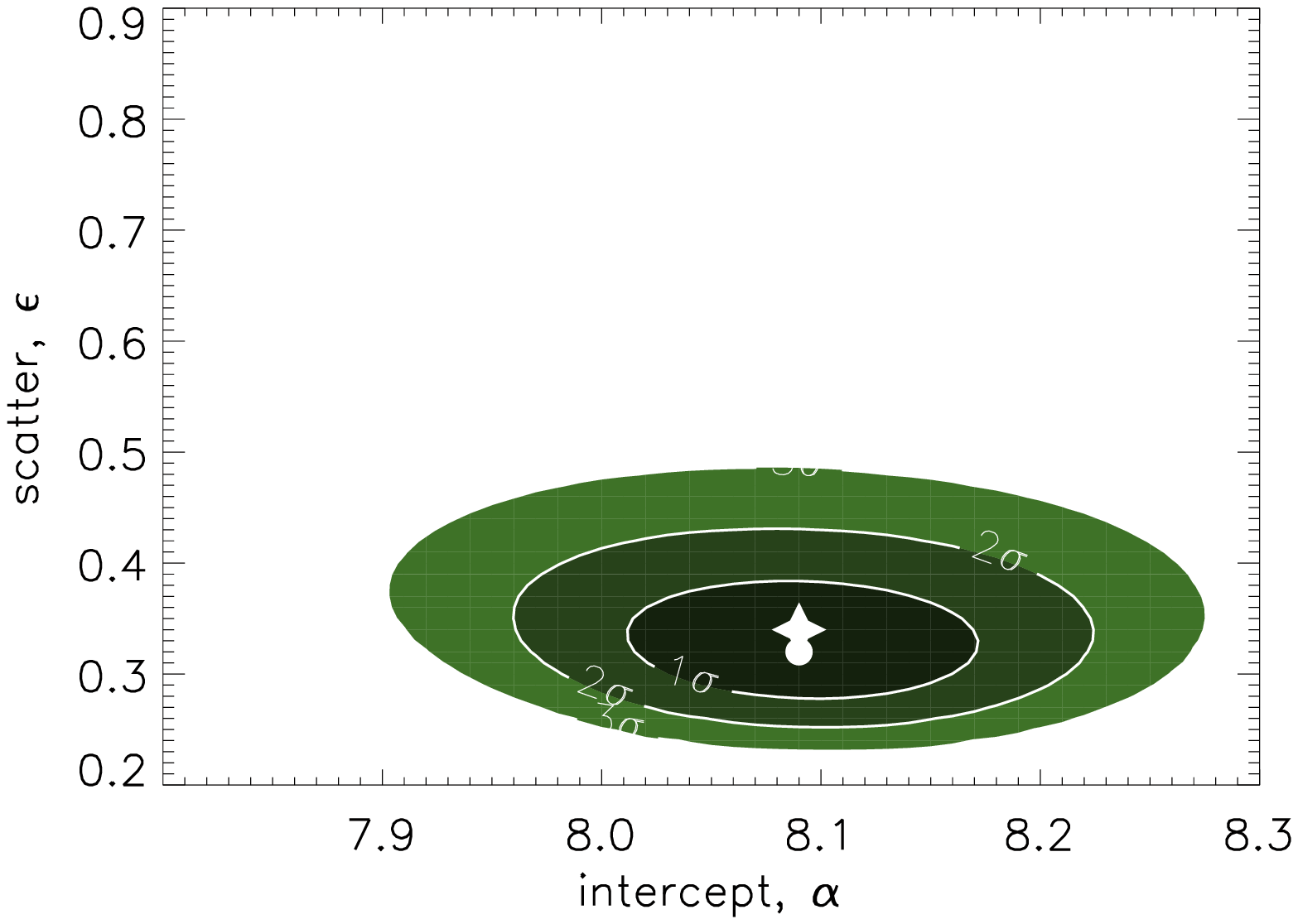}}
\subfigure{\includegraphics[width=0.3\textwidth ,clip]{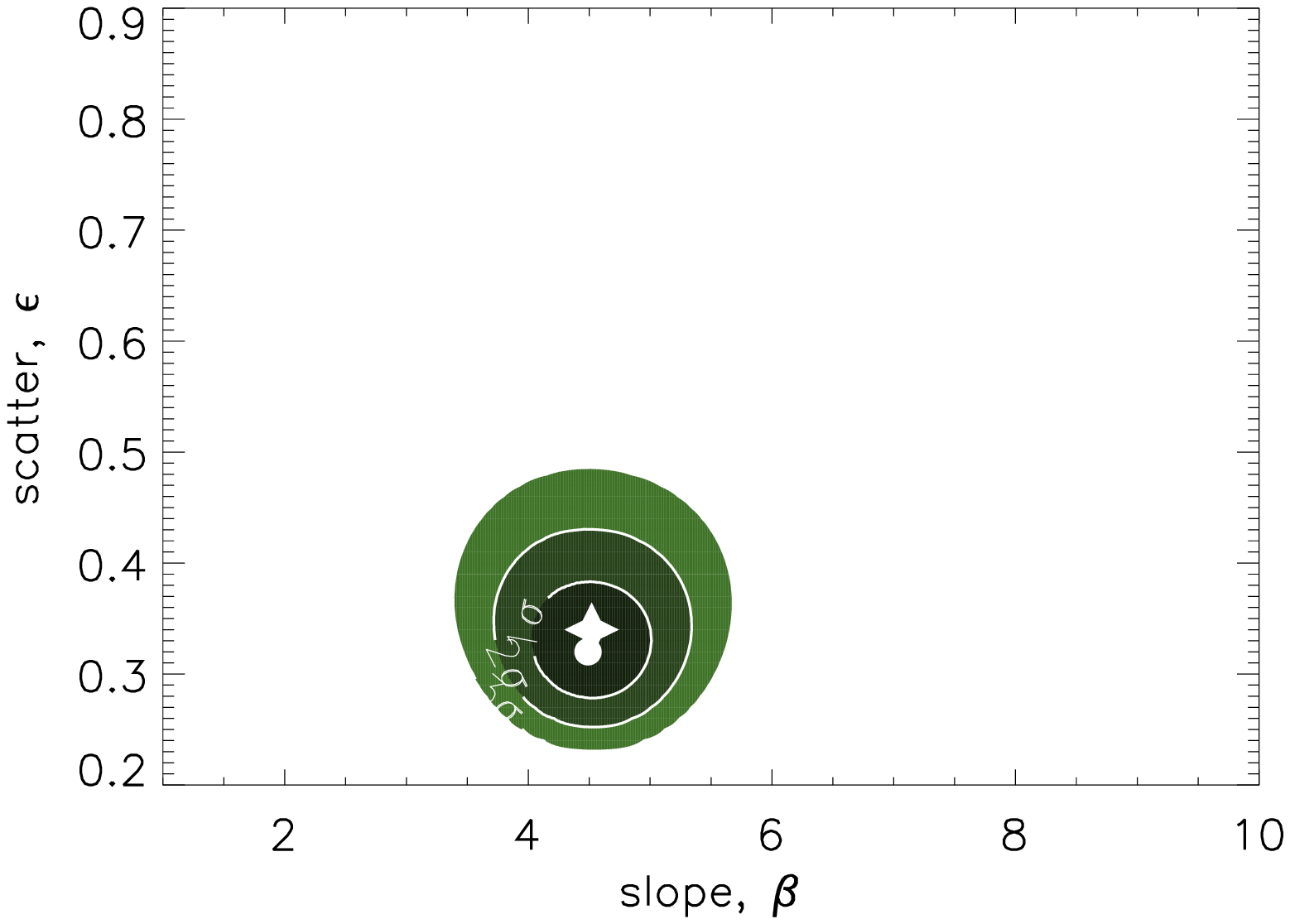}}
\subfigure{\includegraphics[width=0.3\textwidth ,clip]{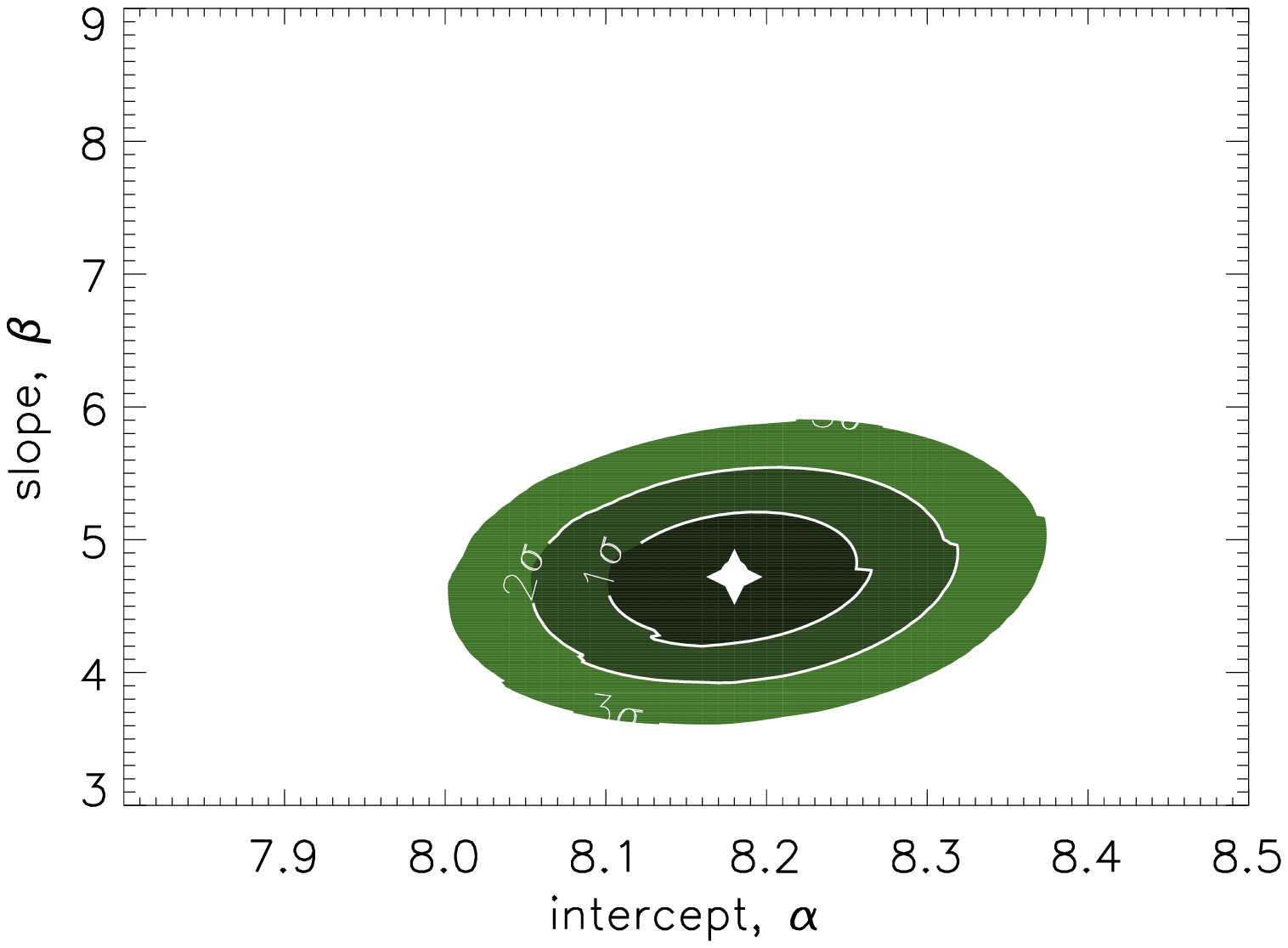}}
\subfigure{\includegraphics[width=0.3\textwidth ,clip]{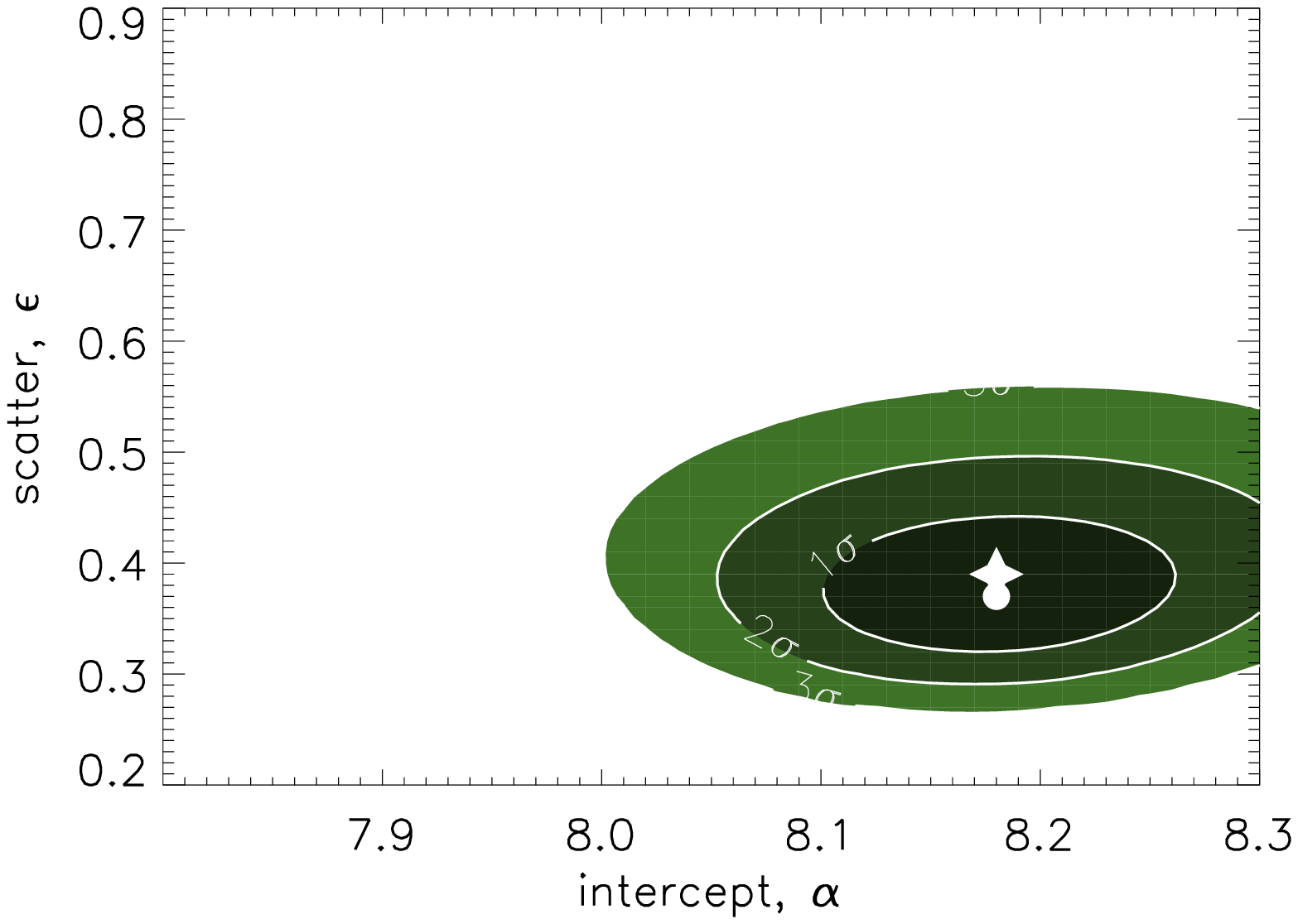}}
\subfigure{\includegraphics[width=0.3\textwidth ,clip]{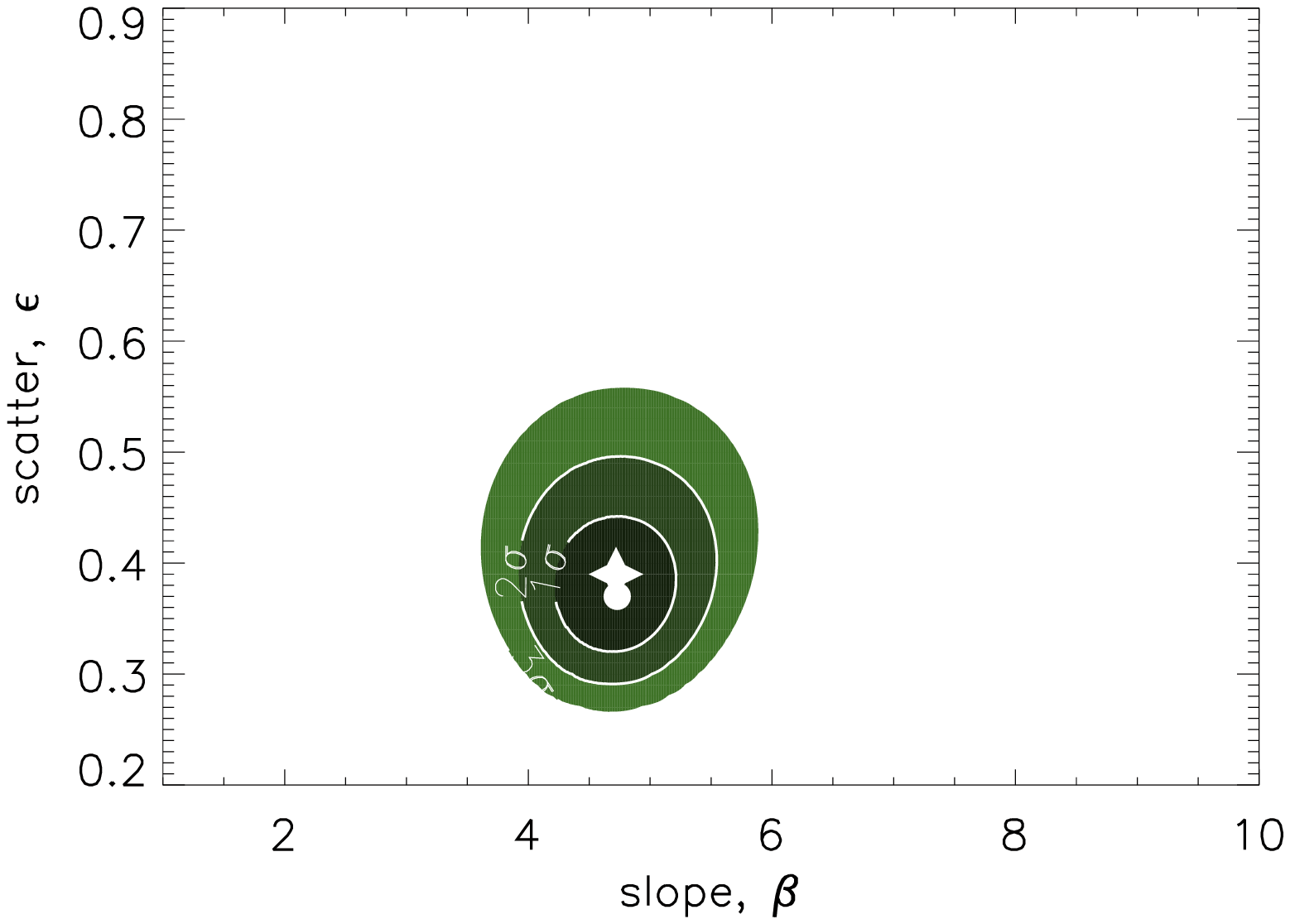}}
\subfigure{\includegraphics[width=0.3\textwidth ,clip]{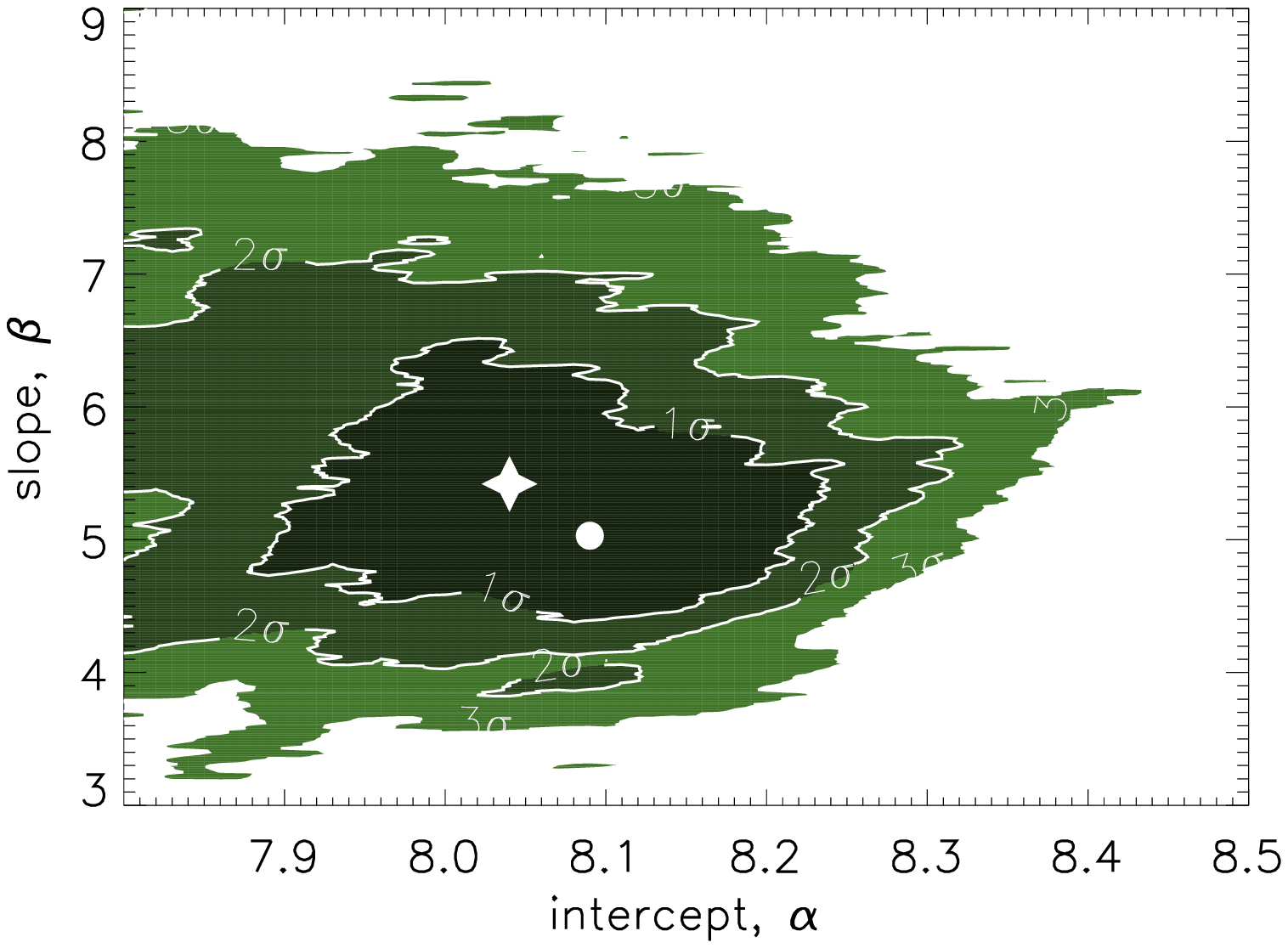}}
\subfigure{\includegraphics[width=0.3\textwidth ,clip]{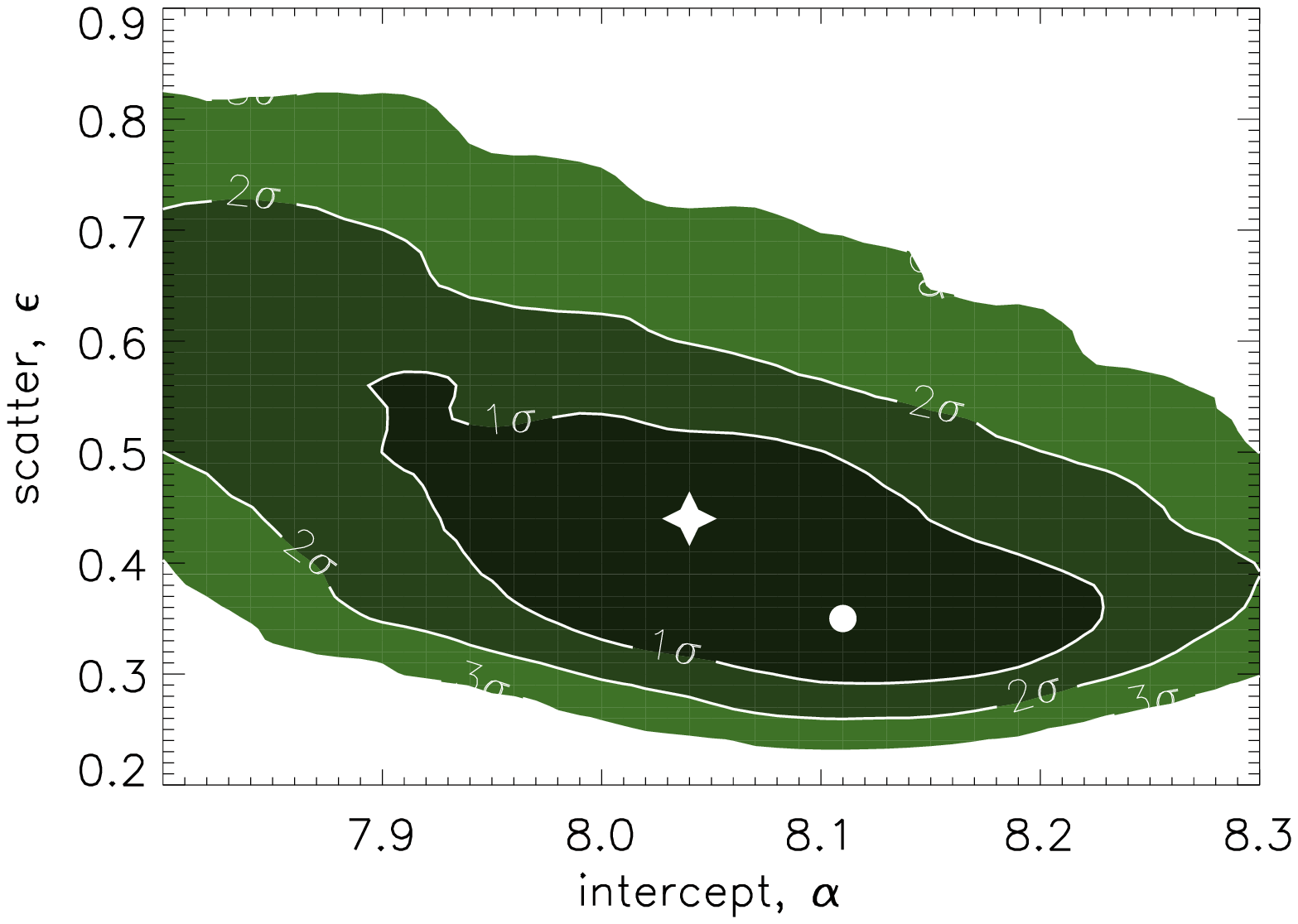}}
\subfigure{\includegraphics[width=0.3\textwidth ,clip]{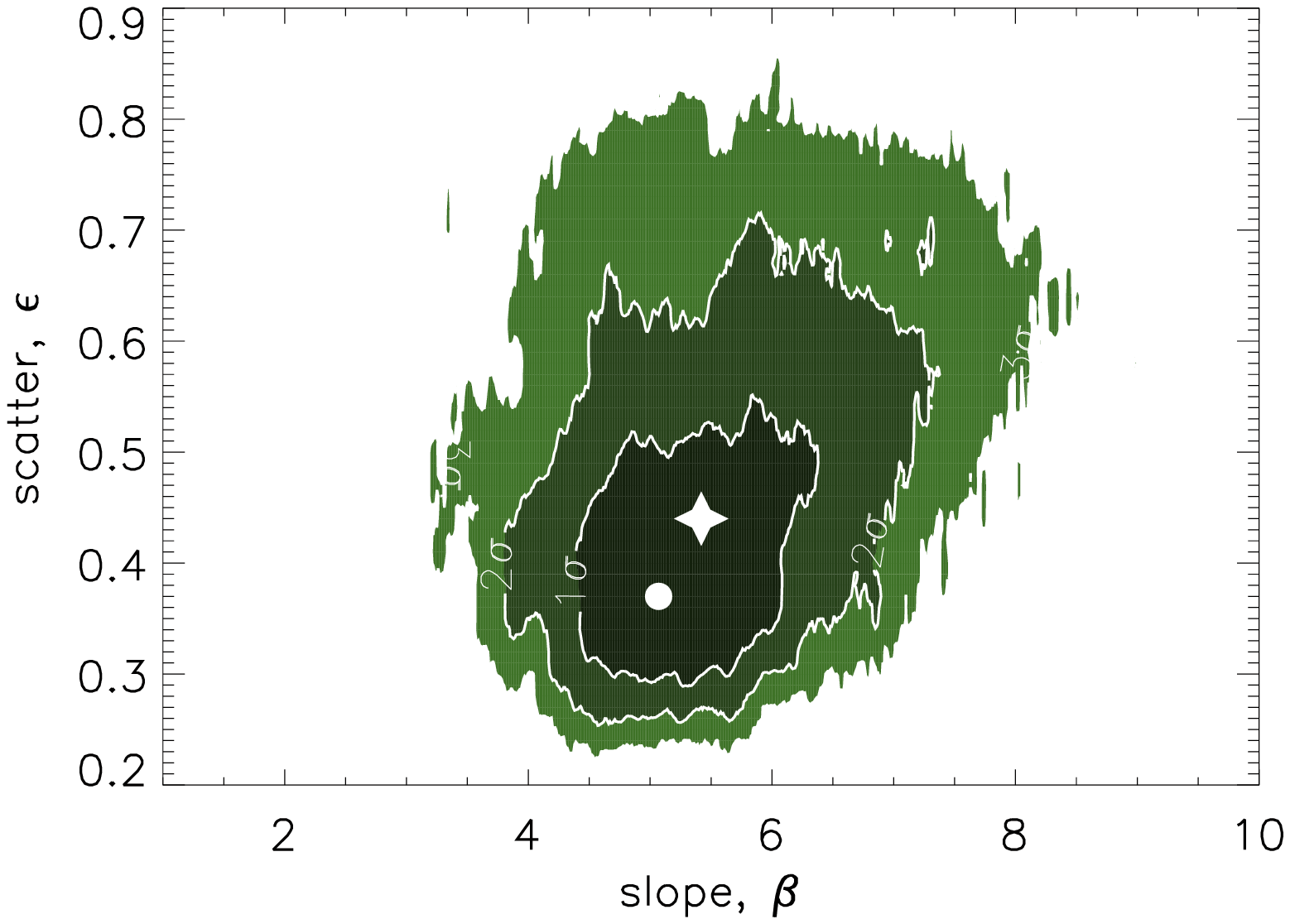}}
\subfigure{\includegraphics[width=0.3\textwidth ,clip]{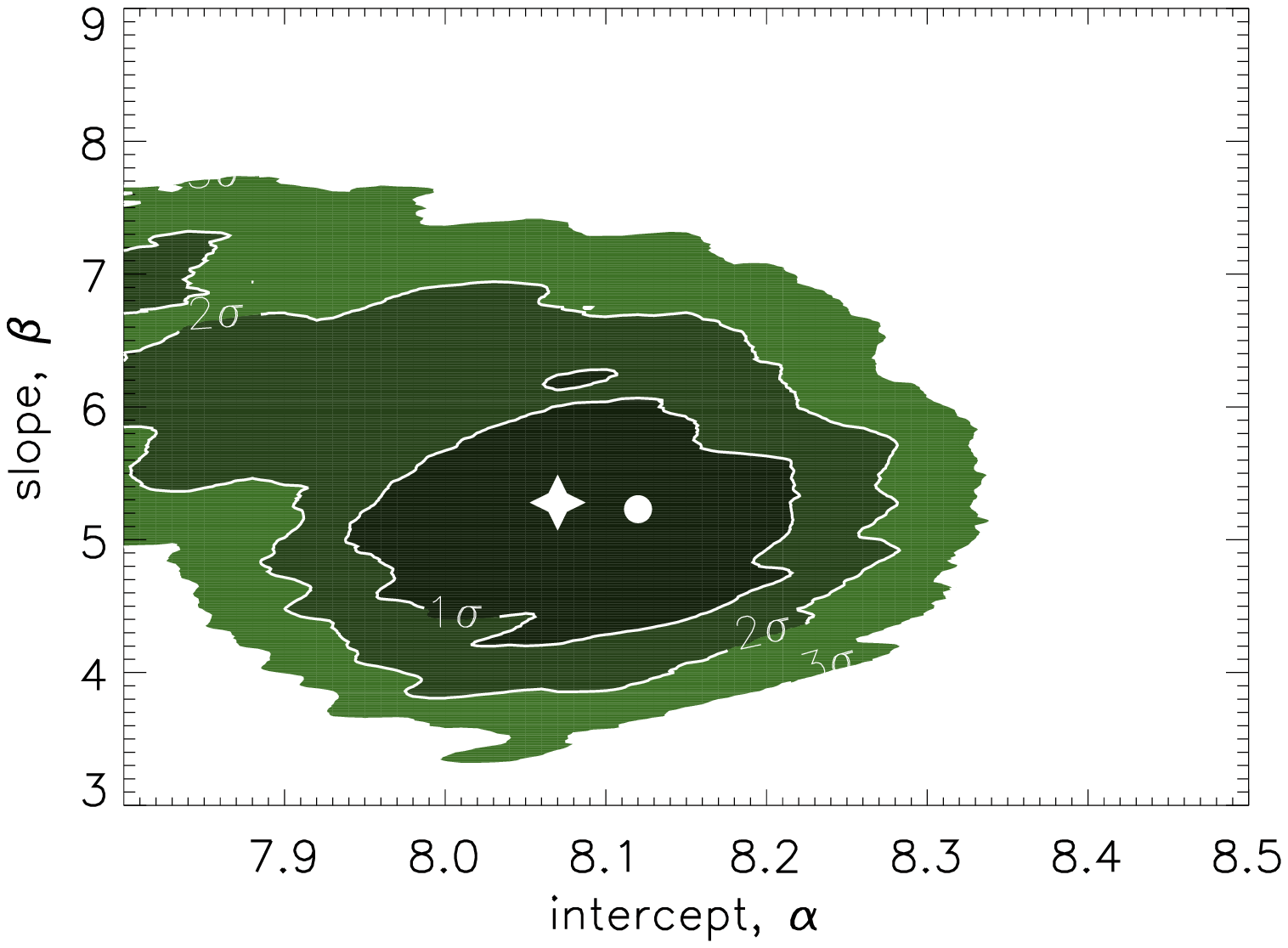}}
\subfigure{\includegraphics[width=0.3\textwidth ,clip]{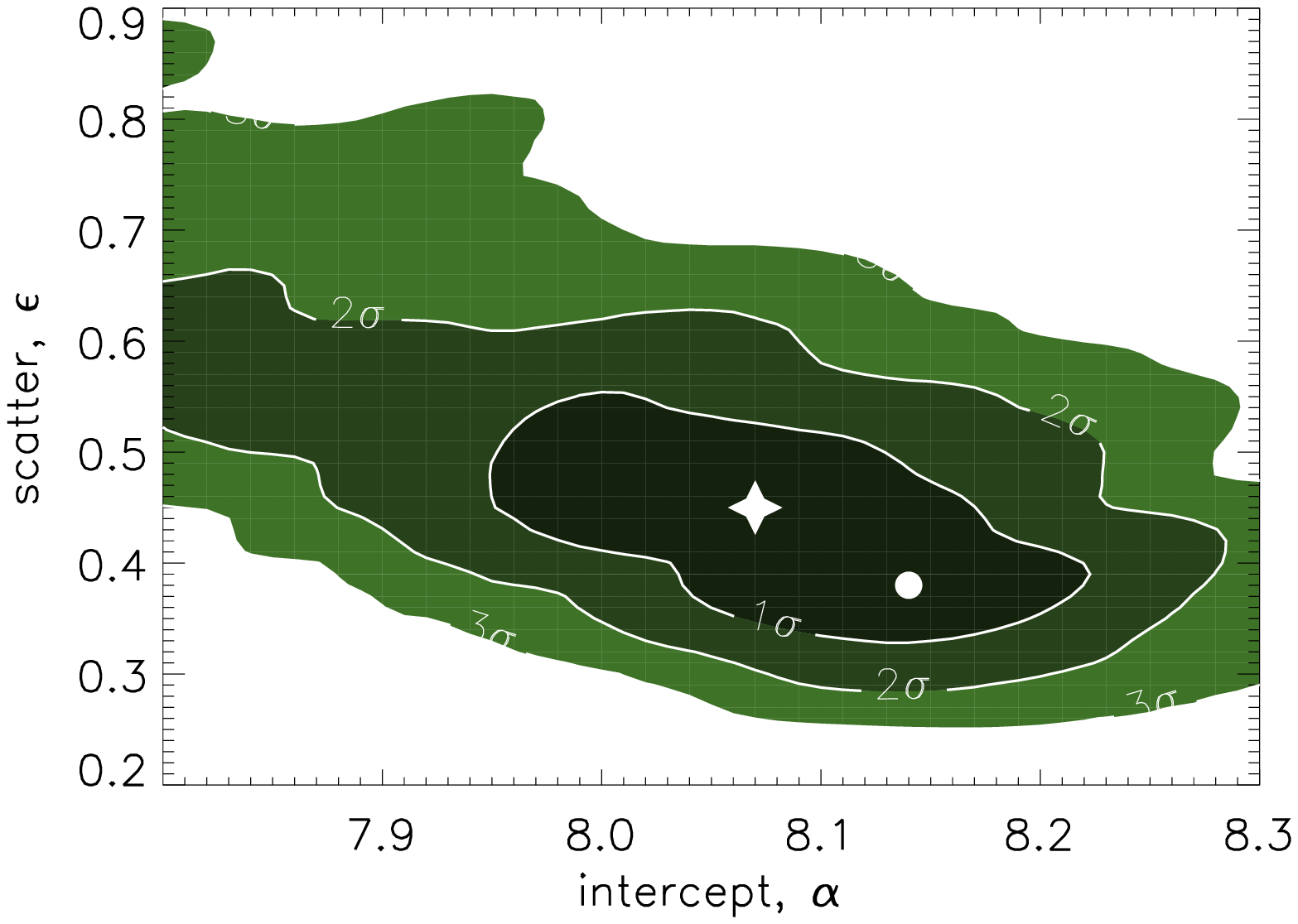}}
\subfigure{\includegraphics[width=0.3\textwidth ,clip]{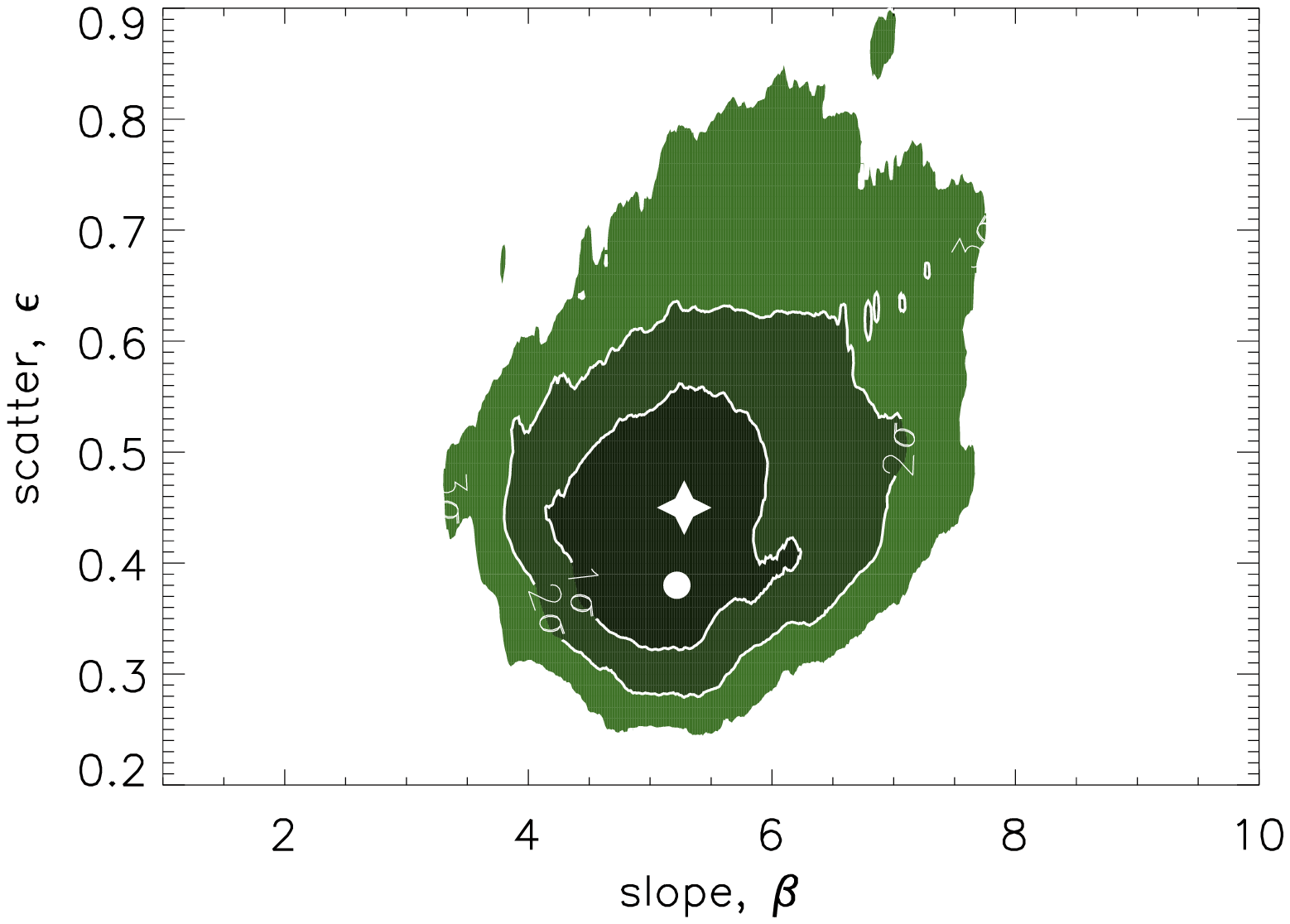}}
\end{center}
\caption[$\chi^2$-likelihood Contours]{The shading represents different confidence levels: the darkest shading shows the 68\% contours, and increases to the 99.73\% limit at the lightest shading. The highest probability in a simulation is marked with a circle, and the expected values are marked with a star. The top row 
displays the contours for the simulation without selection effects, the second row the velocity dispersion selection bias ($\mathcal{S}_{\sigma}$), the third row  shows the simulation with the sphere of influence selection effect ($\mathcal{S}_{r_i}$),  and the bottom row is with both selection effects ($\mathcal{S}_{\sigma,r_i}$) applied. From left to right, the columns show $\beta$ vs. $\alpha$, $\epsilon$ vs. $\alpha$, and $\epsilon$ vs. $\beta$ for the expected value of the third parameter. }
\end{figure}

\subsection{Models (ii) and (iii): No \msig\ Dependency}

It is important to test whether the upper bound of the \msig\ relation can be 
reproduced with a more general model between SMBH mass and bulge velocity dispersion, given the presence of the sphere of influence selection effect. Models (ii) and (iii) assume that the velocity dispersions and SMBH masses are independent variables. We draw the velocity dispersions from HyperLeda, and randomly assign them to simulated SMBH masses. We test two different SMBH distributions. The value of these tests lie not only in seeing if we can reproduce the upper bound of the $M_{bh}-\sigma$ relation with SMBH mass distributions that are independent of $\sigma$, but also as a comparison with Model (i), where an intrinsic relation is assumed.

\subsubsection{Model (ii): Power Law}
The first SMBH mass distribution is described by a power law distribution, $P(M) \propto M^{-\Gamma}$. The power law index ranges from $1\le\Gamma\le1.6$. 
We compare the probability function of this model with the peak probability
 of the linear \msig\ model from our previous simulation, and plot the ratios in the top left plot of Figure~\ref{fig:sims}.
 This ratio is shown for every value of $\Gamma$ within our simulated range, for the base model without applying selection effects and the models with selection functions applied. 
The ratios, therefore, represent the likelihood of the hypothesis that velocity dispersion and SMBH mass are independent variables compared with the hypothesis that SMBH mass depends on velocity dispersion. 
 It is immediately obvious from the top left panel of Figure~\ref{fig:sims} that the selection functions drastically increase the probabilities. There is a significant difference between selection functions that use the sphere of influence selection effect and those that do not; we plot them in individual panels since those without the sphere of influence selection effect applied are several orders of magnitude smaller. We further examine the simulation with selection function $\mathcal{S}_{\sigma,r_i}$ applied, but the simulations are never able to reach the $1\sigma$ point needed in order to be consistent with the GR11 sample (depicted in the first row of Figure~\ref{fig:sims}, along with the expected value of the power law slope, $\bar\Gamma=1.45_{-0.05}^{+0.50}$). The bottom panel of Figure~\ref{fig:sims} plots the simulation for $\bar\Gamma=1.45$ along with the GR11 sample.  There are too many outliers in the simulation above the relation, and too few points at the low velocity dispersion end of the simulation. This does not accurately capture the behavior of the observed \msig\ relation, and can therefore rule out the model of a power law mass distribution coupled with selection effects as an explanation of the empirical \msig\ relation.

\begin{figure}[H]
\figurenum{7}
\begin{center}
\subfigure{\includegraphics[width=0.4\textwidth ,clip]{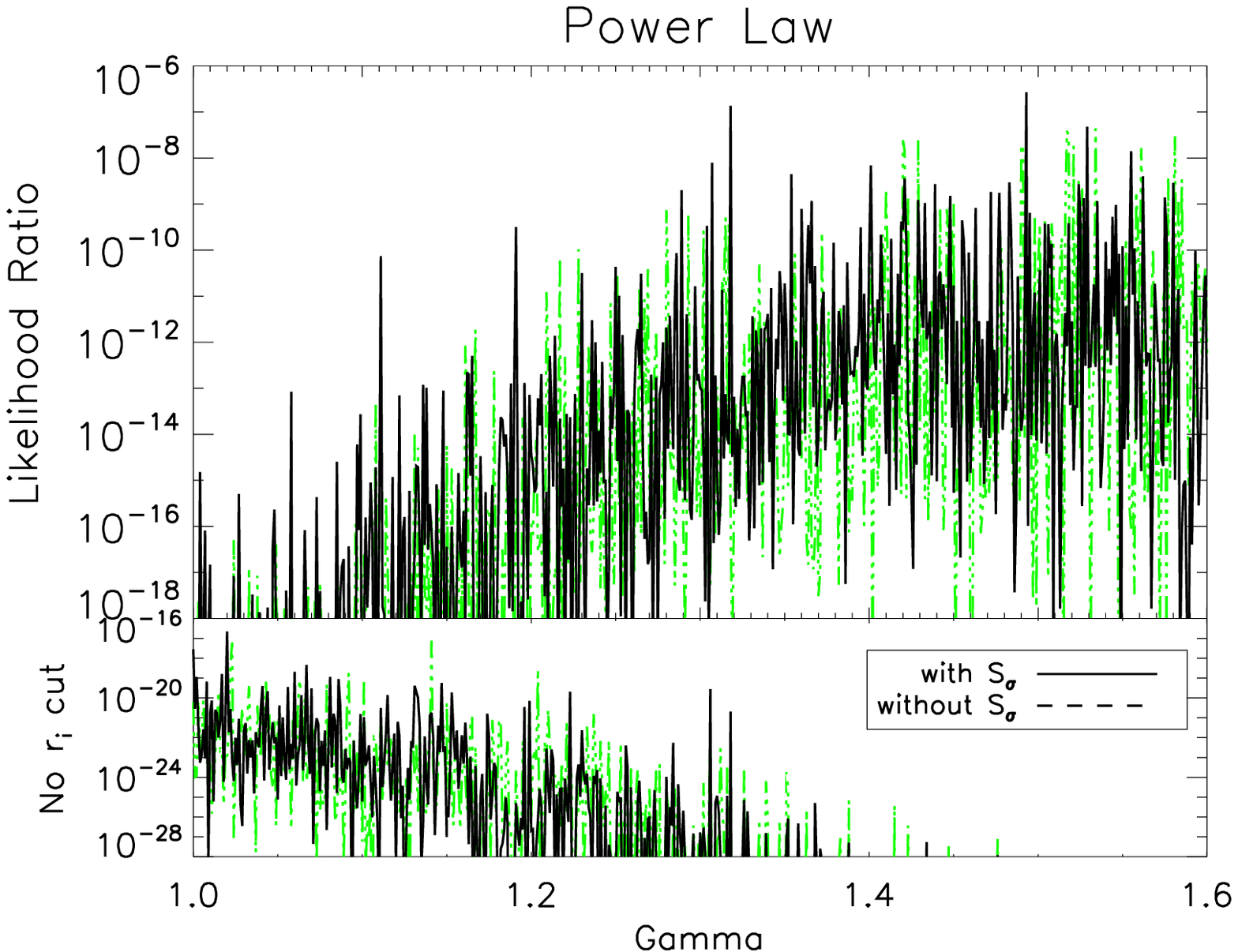}}
\subfigure{\includegraphics[width=0.4\textwidth ,clip]{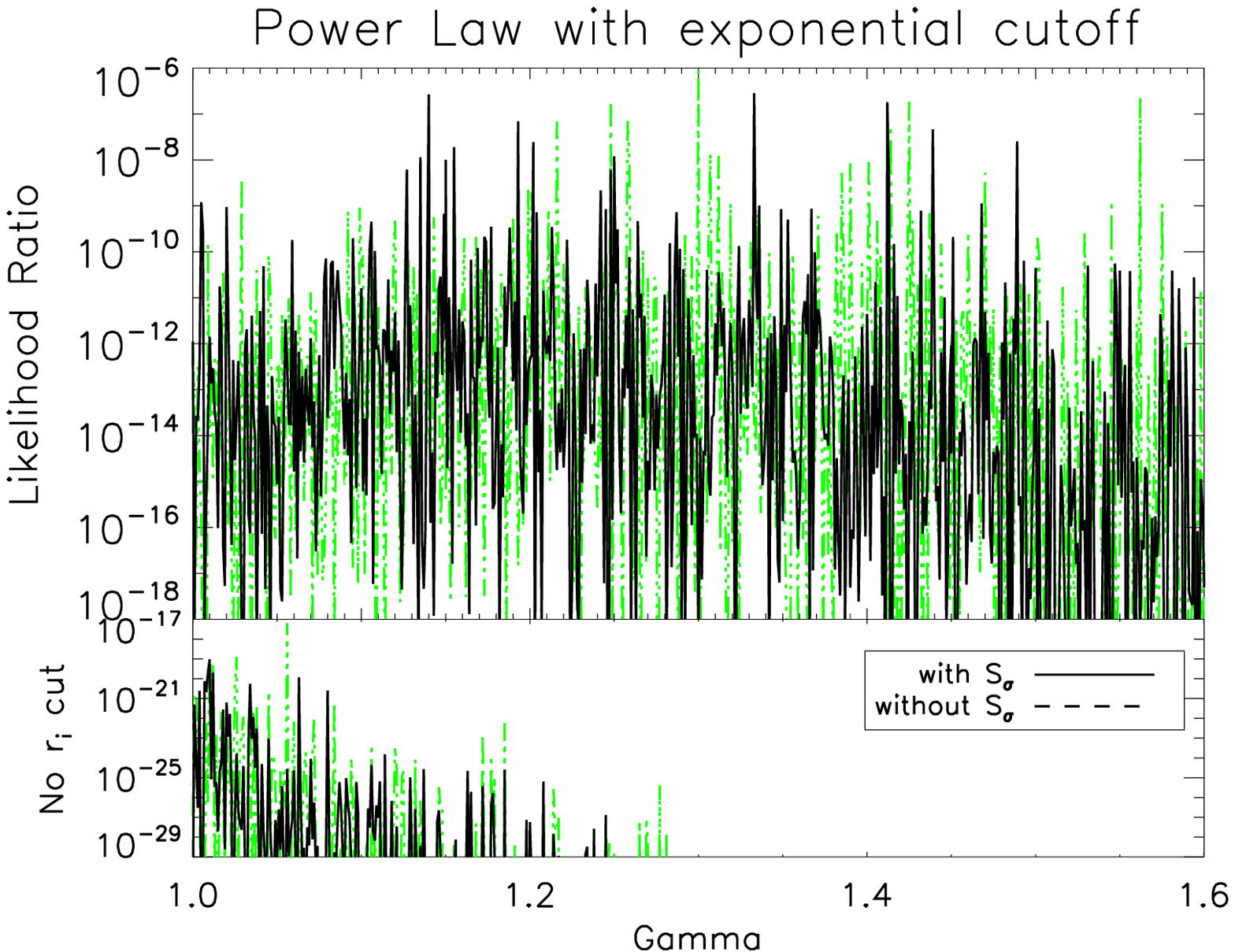}}
\subfigure{\includegraphics[width=0.4\textwidth ,clip]{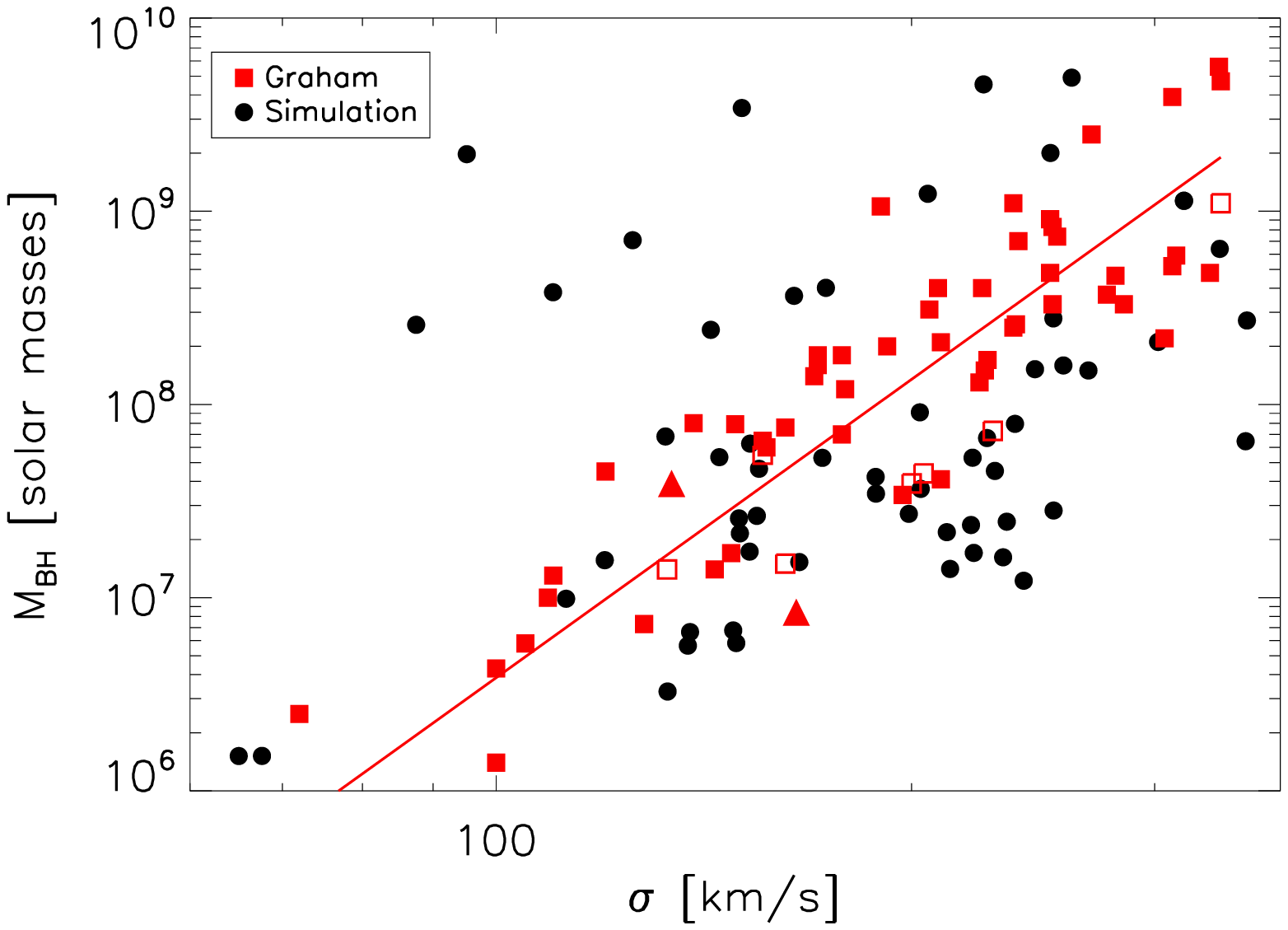}}
\subfigure{\includegraphics[width=0.4\textwidth ,clip]{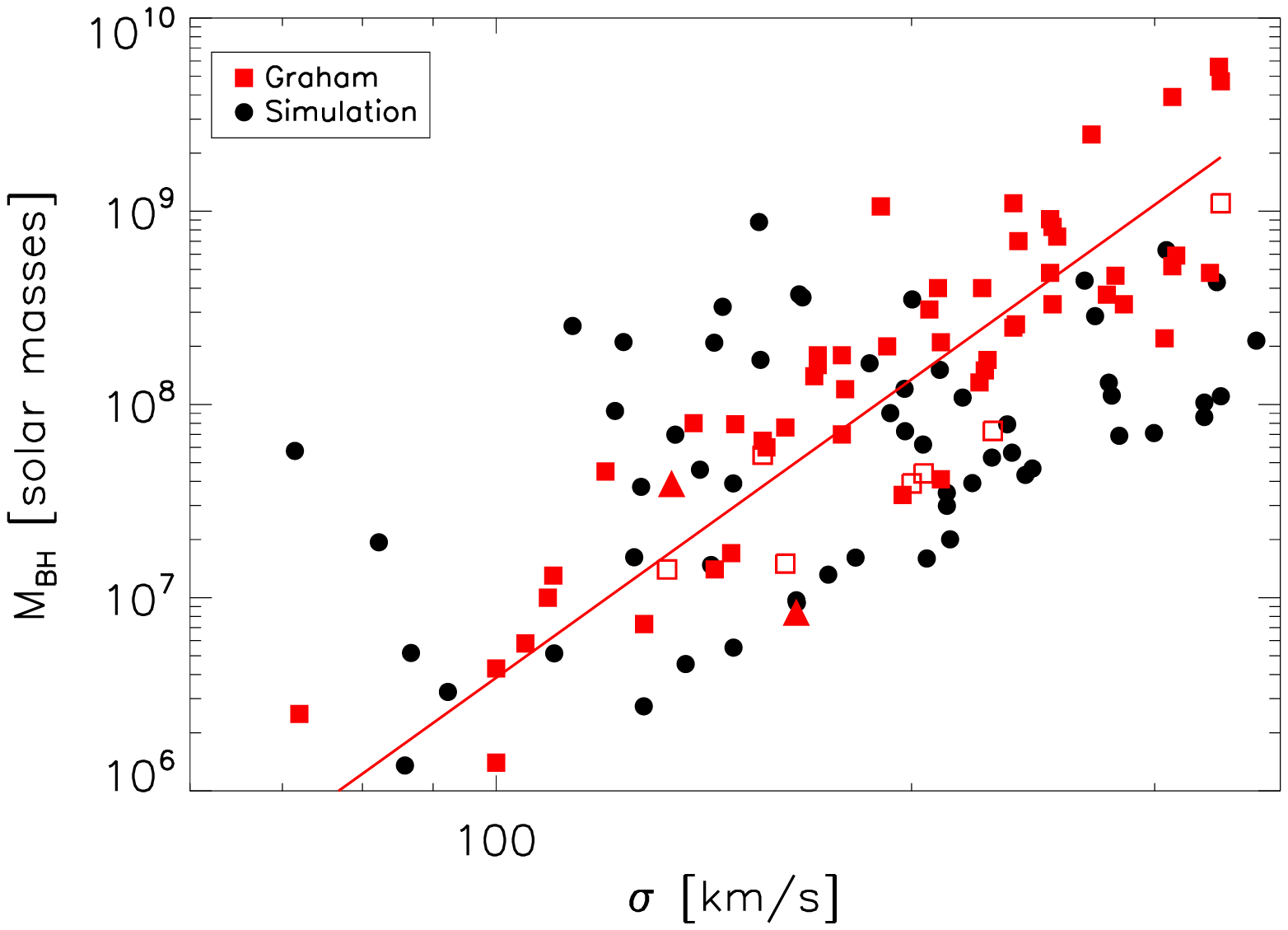}}
\end{center}
\caption[Power Law and Exponential Cutoff Mass Distributions]{For power law (left) and exponential cutoff (right) mass distributions: the top panels show the $\chi^2$-likelihood ratio of the power law or exponential cutoff distribution to the maximum $\chi^2$-likelihood value for the linear \msig\ relation simulation, plotted as a dashed green line. The top panels have the sphere of influence selection effects applied. The bottom plots show the GR11 sample overlaid with the $\mathcal{H,S}_{\sigma,r_i}$ simulation for the expected values $\bar\Gamma=1.45$ (power law) and $\bar\Gamma=1.30$ (power law with exponential cutoff). 
}
\label{fig:sims}
\end{figure}

\subsubsection{Model (iii): Exponential Cutoff}
The last case considered is a power law distribution with an exponential cutoff. This is the same as Model (ii) but with an exponential cutoff set at $10^9$\msun. The cutoff shifts the expected value to $\bar\Gamma=1.30_{-0.14}^{+0.13}$ for the simulation with $\mathcal{S}_{\sigma,r_i}$ applied. The right side of Figure~\ref{fig:sims} shows the results for the expected value. This model is also not able to reach a point where it is consistent with the GR11 sample. The clustering of simulated galaxies is too heavy on the low-mass side of the relation, and while the scatter is tighter than in the power law model, there are still many outliers in this area, and not enough in the high mass end of the distribution. Therefore, we can also rule out this SMBH mass distribution combined with selection effects as an explanation for the \msig\ relation.

\section{Discussion and Conclusions}
Using a Bayesian Monte-Carlo analysis, we constrain the parameters of the \msig\ relation as $\alpha = 8.07_{-0.10}^{+0.08}$, $\beta = 5.28_{-0.55}^{+0.84}$, and $\epsilon = 0.45_{-0.07}^{+0.08}$. The main results of this paper are:
(1) the directly measured slope of an \msig\ relation is shallower than the intrinsic slope because of selection effects in \ri\ and \vd , and (2) the uncertainties on the slope of the relation are likely underestimated in previous measurements.  We additionally conclude that the \msig\ relation cannot be successfully reproduced with either power law or exponential cutoff mass distributions coupled with \ri\ and \vd\ selection effects, and that it is statistically much more likely that the relation is intrinsic rather than observed.

Comparing the two selection effects shows clearly that the sphere of influence selection effect dominates, which will decrease the slope if the intrinsic slope is steep, e.g., $\beta \sim 10$, or increase the slope if the intrinsic slope is shallow, e.g., $\beta \sim 1$, with the crossing point at $\beta \sim 4$ (Figure~\ref{fig:inout}).
The \ri\ selection resembles a parabolic curve in the logarithmic \msig\ plane, which can be approximated with a linear cut of $\beta \sim 3.5$. For a steep intrinsic \msig\ relation, more objects with lower masses will be cut if one applies the \ri\ resolution selection criteria, resulting a shallower relation. 
 The situation will reverse for very shallow intrinsic slopes (e.g., $\beta \sim 1$), where the simulated slope will be steeper than the intrinsic input slope after applying selection effects. However, since the likelihood function for these cases approaches zero, this effect is not important. 
 For the power law and exponential cutoff mass distributions, our models assume more galaxies with smaller \m\ given the range of $\Gamma$ used in the simulations.  However, in the high \vd\ range, the \ri\ selection will trim most of the low mass objects, and in the low \vd\ range, only a few objects have measurements in \m\ and it is more likely for the mass to be small based on our assumptions.
Even though we are able to rule out these two scenarios, it is still clear that selection effects can significantly increase the chance that these intrinsic power law models would be consistent with the observed sample.

The \vd\ selection effect can be important if the intrinsic relation between \m\ and \vd\ is non-linear as in our power law models.  For linear models, the \vd\ selection effect can also be important if the intrinsic slope is extremely deep, $\beta \sim 10$, or shallow, $\beta \sim 1$, where the effect can modify the slope by $\Delta \beta \sim 1$.  Fortunately, within the range of intrinsic slopes of $\beta = 4$--6, the \vd\ selection effect is not important.

The argument for using an \ri\ selection criteria is that black hole mass estimates can be unreliable \citep[e.g.,][]{ff05}. Ensuring that the sphere of influence is resolved is thought to more accurately measure the \msig\ relation. 
In this paper, we find that this will cause selection effects that impact estimating the best-fit parameters, especially the slope $\beta$.  Based on our simulations, we find
\begin{equation}
    \beta_{intr} = -4.69 + 2.22\times\beta_{measured},
\end{equation}
where $\beta_{intr}$ is the intrinsic slope and $\beta_{measured}$ is the directly measured slope without modeling the selection effects.
In the relevant parameter space $\beta = 4$--6, the \ri\ selection effect decreases the measured slope.
This is consistent with some previous results \citep[e.g.,][]{fm00}. The authors measured the \msig\ relation twice, once with secure measurements and the sphere of influence resolved, and once for their entire sample. They found that the slope decreases from $5.81\pm0.43$ to $4.80\pm0.54$ when including only galaxies where \ri\ was resolved. 
\citet{gultekin09} used Monte-Carlo simulations to show that an \ri\ selection criteria can bias measurements of the zero point, slope, and scatter. They consider three different scenarios using synthetic data sets. In the first scenario, they require that \ri\ be resolved based on the values of $M_{bh}$ and \vd\ in the data set. For the next scenario, they use \vd\ to calculate the expected $M_{bh}$, find the radius of the sphere of influence using \vd\ and the expected mass, and then cut any data pairs where \ri\ is not resolved. The last scenario uses an observed sample of \vd\ with simulated masses. The results here agree with their first scenario, as the slope decreases with selection effects. However, we find an opposite bias in the slope for the third scenario, which is closest to the simulations here as it uses observed \vd. 
The differences might stem from the fact that the \citet{gultekin09} scenario uses observed distances and instrumental resolution, which give different values for \ri, rather than assuming a standard resolution for all data points, as done here.
In addition, \citet{gultekin09} studies a sample yielding a low slope, which, based on our simulation, has a smaller correction on $\beta$ (Figure~\ref{fig:inout} and Equation~2).
To test the robustness of our conclusions, we run our Bayesian Monte-Carlo code with other recent samples \citep{gultekin09,mc11, beifiori12}.
We list our corrections on the slope, $\Delta \beta$, in Table~\ref{slopes}.
We find decreases of the measured slope after applying selection effects in all the samples, however, with larger corrections for larger measured slopes.  This is consistent with the results of our simulation. 

There are a number of measurements of the \msig\ relation, but we do not find an analysis that fully models various selection effects. The fitting results among  different groups are not consistent, due to the mass measurements, sample selections, or linear regression fitting techniques (about $\Delta \beta \sim 0.2$). 
Using the sample of \citet{graham11} but filtering out the two galaxies not satisfying our simulation resolution of 0\sarc08, we find a best estimated slope of $\beta = 5.28_{-0.55}^{+0.84}$ by modeling the selection effects. 
Although this is in agreement with studies that measure higher values for the slope (e.g., \citet{ff05}, $\beta=5.81\pm0.43$; \citet{graham11}, $\beta=5.13\pm0.34$; \citet{gr08}, $\beta=5.52\pm0.40$), this consistency needs to be further evaluated.
For example, our analysis indicates that directly fitting the \citet{graham11} sample should yield a slope of $\beta \sim 4.5$ (Table~2), and indeed, we find a slope of $\beta \sim 4.6$ by directly performing a linear regression on the volume-limited \citet{graham11} sample with which we compare our simulations.
Applying the \ri\ selection is important to keep the mass measurement reliable; however, we also need to model this effect to restore the intrinsic slope of the relation.
For groups measuring a shallower slope, $\beta \sim 4$, the selection effect is small based on our simulation, and for measurements of steeper slopes $\beta \sim 5$, the selection effect is larger.
Therefore, it appears the difference in the measurements of slopes ($\sim 4$ vs. $\sim 5$) among different groups arises mainly from the basic mass measurements.  

An \msig\ relation has far reaching implications, as the slope of the relation can illuminate important physical processes connecting a central black hole to the host galaxy. The results presented here suggest that the slope is likely higher than is measured due to \ri\ and \vd\ selection effects, and closer to the energy-driven winds scheme as predicted by \citet[$M\propto\sigma^5$;][]{silk98} rather than the momentum-driven winds predicted by \citet[$M\propto\sigma^4$;][]{fabian99}. 
This feedback mechanism can be realized by quasar winds, especially in broad absorption line quasars (BALQSOs). The efficiency depends linearly on the covering fraction of BALQSOs, and recent studies suggest that the intrinsic fraction of BALQSOs in quasars is at least two times higher than those fractions measured in optical surveys \citep[e.g.,][]{dai08,s08,allen11}.  For low-ionization BALQSOs (LoBALs),  including ones exhibiting iron absorption (FeLoBALs), the intrinsic fractions are estimated to be 5--7 times larger than the values obtained from optical surveys \citep{dai10b}.  Although the intrinsic fractions of LoBALs are still small, the column density in these wind can be orders of magnitude higher than other quasar winds.  Using more robust measurements of the column densities of FeLoBALs, it is estimated that FeLoBALs can provide kinetic feedback efficiency $\dot{E}_{kin}/L_{bol}$ of a few percent \citep[e.g.,][]{m11}, consistent with the requirement for the feedback efficiency of the \msig\ relation \citep[e.g.,][]{silk98}.
Although the energy feedback model can suffer from radiation loss \citep[e.g.,][]{k03,sn10}, it is possible that the feedback process is a multi-phase process, where the AGN feedback is just the first phase \citep[e.g.,][]{he10,sn10}.  Subsequent feedback from supernovae and stellar winds will provide additional energy.  The slope of the predicted \msig\ relation is still $\propto \sigma^5$ if the feedback is dominated by energy exchanges \citep{sn10}.

An alternative model that is not studied here is the combination of an upper envelope and selection effects causing us to see only the upper portion of the \msig\ plane \citep{b10}, as physically interpreted by \citet{king10}. However, \citet{gultekin11} argue that these upper-limit models predict detection rates of black hole masses that are not observed. The lower portion of the \msig\ plane will be probed as technology and techniques advance
enough to be able to resolve low mass black holes. Continuing to expand the observed
samples with reliable SMBH mass estimates is essential, but until we reach the point where
selection effects are negligible, they must be well modeled when analyzing a sample.

Here we have only tested a limited number of hypotheses against the current data, and
therefore our conclusions are limited to those hypotheses. There are other factors that are not
explored: for example, many studies differentiate galaxies by morphology when measuring
the \msig\ relation (e.g., Graham et al. 2011; Hu 2008; Beifiori et al. 2012). While there are
tantalizing hints that the \msig\ relation may be different for different galaxy populations,
the sample sizes are still too limited to be conclusive. This paper provides a formalism that
is capable of testing a large number of hypotheses combined with selection effects, a useful
springboard for future, deeper studies. 

\acknowledgements
We acknowledge the use of the HyperLeda database (\url{http://leda.univ-lyon1.fr}). We would also like to thank the anonymous referee for helpful suggestions.

\end{document}